\documentclass{aa}  
\bibpunct{(}{)}{;}{a}{}{,}
\usepackage{graphicx}
\usepackage{txfonts}
\usepackage{indentfirst}
\usepackage{tablefootnote}
\authorrunning{N. L. Rossignoli et al.}
\usepackage{dblfloatfix}
\usepackage{cuted}
\usepackage{hyperref}
\usepackage{xcolor}
\definecolor{Azul1}{rgb}{0,0.086,0.6}
\hypersetup{
       colorlinks=true,
       citecolor={black},
       linkcolor={Azul1},
       urlcolor={Azul1},
   }
        
\begin{document}

   \title{Cratering and age of the small Saturnian satellites}


   \author{N. L. Rossignoli
          \inst{2},
          R. P. Di Sisto
          \inst{1,2},
          M. Zanardi
          \inst{2}
          \and
          A. Dugaro
          \inst{1,2}
          }
   \institute{Facultad de Ciencias Astronómicas y Geofísicas, Universidad Nacional de La Plata, Paseo del Bosque S/N (1900), La Plata, Argentina
         \and
             Instituto de Astrofísica de La Plata, CCT La Plata - CONICET - UNLP, Paseo del Bosque S/N (1900), La Plata, Argentina}
   \date{}

  \abstract
{The small ($\le$ 135 km mean radius) satellites of Saturn are closely related to its rings and together they constitute a complex dynamical system where formation and destruction mechanisms compete against each other. The Cassini-Huygens mission provided high-resolution images of the surfaces of these satellites and therefore allowed for the calculation of observational crater counts.}
{We model the cratering process by Centaur objects on the small Saturnian satellites, and compare our results with the observational crater counts obtained from the Voyager and Cassini missions.}
{Using a theoretical model previously developed we calculate the crater production on these satellites considering two slopes of the size-frequency distribution (SFD) for the smaller objects of the Centaur population and compare our results with the available observations. In addition, we consider the case of catastrophic collisions between these satellites and Centaur objects and calculate the age of formation of those satellites that suffer one or more disruptions.} 
{In general we find that the observed crater distributions are best modeled by the crater size distribution corresponding to the $s_2 = 3.5$ index of the SFD of impactors with diameters smaller than 60 km. However, for crater diameters $D \lesssim 3-8$ km (which correspond to impactor diameters  $d \sim 0.04 - 0.15$ km), the observed distributions become flatter and deviate from our results, which may evidence processes of erosion and/or crater saturation at small crater sizes or a possible break in the SFD of impactors at $d \sim 0.04 - 0.15$ km to a much shallower differential slope of $\sim -1.5$. Our results suggest that Pan, Daphnis, Atlas, Aegaeon, Methone, Anthe, Pallene, Calypso, and Polydeuces suffered one or more catastrophic collisions over the age of the solar system, the younger being associated to arcs with ages of $\sim10^8$ years. We have also calculated surface ages for the satellites, which indicate ongoing resurfacing processes.}
{}
 
   
  
  
   \maketitle
%

\section{Introduction}
   \label{intro}
Craters on solar system objects generated by collisions represent one of the most distinctive marks on their surfaces, allowing us to peek into their past to discover their history and evolution. Extended missions to the solar system objects have enabled us to study these structures in great detail and to pose new questions about the physical, dynamical, chemical, and even biological processes that take place in them. More recently, the Saturn system has been studied in depth by the Cassini-Huygens mission, which made numerous and impressive discoveries: Titan and Enceladus were found to harbor complex chemical environments of astrobiological interest, diverse features and processes were observed in the constantly changing ring system, and many tiny moons were uncovered, which we included in the present study.

The small ($\le$ 135 km mean radius) satellites of Saturn, with the exception of Hyperion, are closely related to the rings and together they constitute a dynamical system where a variety of physical processes take place and compete against each other \citep{Porco2007}. Formation and destruction mechanisms are evident when looking at the whole scenario of satellites plus rings. The current situation of the satellites, their existence, and their surface properties are the reflection of what happened during the formation of the solar
system, that is, the primordial scenario and its subsequent dynamical and physical evolution. Thus, the formation of the small Saturnian satellites and their eventual destruction by tidal effects or by collisions as well as the existence of rings all seem to be aspects of the same matter \citep{Charnoz2018}. \citet{Charnoz2010} demonstrated that the small Saturnian moons could have formed by gravitational collapse from the ring material around the planet. They obtained mass distributions of moonlets that accurately matched the observed distribution. However, the formation of rings and satellites in the giant planets is not a closed topic and different solutions have been considered and studied \citep{Canup2010, Crida2012, Hyodo2017, Salmon2017}.   

Nevertheless, the small moons studied in the present work, with the exception of Hyperion, are associated with the rings. Either orbiting Saturn at the edge of its rings or embedded in them, the  surfaces of these satellites are altered by their entourage, with some of them showing evidence of erosion and mass transport due to ring particle deposition \citep{Hirata2014, Thomas2013}.

After Voyager observations, the analysis of the images suggested the existence of two types of impactor populations: Population 1, which comprises heliocentric objects producing large craters, and Population II, associated with planetocentric debris which produces small craters \citep{Smith1982}. There are a number of papers on the possible origin and fate of a planetocentric population of debris (e.g., \cite{Marchi2001, Dobrovolskis2004}).
However, there is not a quantitative study about a possible source of craters from a planetocentric source, and therefore it is not possible to calculate their ultimate real contribution. 

In the current solar
system, the small heliocentric bodies that are potential projectiles to produce craters on the satellites of the giant planets are the Centaurs originated in the transneptunian zone. This region beyond Neptune has a well-defined structure divided in four different sub-populations: the classical Kuiper Belt objects (CKBOs) with semi-major axis between $\sim 40$ au and $\sim 50$ au and low-eccentricity and inclination orbits, the resonant objects in mean-motion resonances (MMRs) with Neptune, such as the plutinos in $2:3$ MMR, the scattered disk objects (SDOs) with perihelion distances of $30$ au $ < q < 39$ au that can cross the orbit of Neptune and eventually evolve into the planetary region becoming a Centaur, and the extended scattered disk objects (ESDOs) with $q > 39$ au that are decoupled from Neptune. It is generally accepted that the main source of Centaurs is the scattered disk (SD), since other sources such as plutinos \citep{Morbidelli1997, DiSisto2010}, Neptune Trojans \citep{Horner2010}, or CKBOs \citep{Levison1997} are secondary. 
   
\citet{DiSisto2011} and \citet{DiSisto2013} developed a theoretical model to calculate the cratering of the mid-sized Saturnian satellites produced by Centaur objects. The comparison between their theoretical production of craters, which is independent of the geological processes of a given surface, with the observed crater counts, which are highly affected by the geological history of the object, made it possible to estimate what is known as the ``age'' of the surface \citep{DiSisto2016}. 

In this work we study the small Saturnian satellites: those whose orbits are related to the A ring, those associated to the F ring, the co-orbitals Janus and Epimetheus, the tiny satellites that orbit embedded in arcs between Mimas and Enceladus, Dione and Tethys' trojans, and the highly cratered and chaotic Hyperion. We describe all of these in detail in the following section.
Using a theoretical model that we previously developed, we calculate the crater production on the different satellites generated by Centaur objects. We then compare our results with the crater counts obtained from the Cassini and Voyager observations. This enables us to study the physical consequences of crater production on the small satellites and to establish restrictions on their origin, formation, surface age, and properties.

\section{The small satellites}
The small icy satellites of Saturn present a broad variety of shapes and have radii between 300 m and 135 km. They have high porosities and very low densities, about half the density of water ice.  
Physical and dynamical parameters are shown in Table \ref{propsat}.
Except for Hyperion, they are strongly related to the ring system, some of them being embedded in them and others producing a gap or just orbiting at the border of a ring. Others are related even to the mid-sized satellites as trojans of Tethys or Dione.

The Cassini mission obtained numerous observations of these bodies, enabling the discovery of intriguing physical and dynamical features unprecedented in the solar
system.
 
Next we describe the studied satellites ordered in increasing distance to Saturn.   

\begin{flushleft}
\begin{center}
\begin{table}[h]
\caption[Properties of the small saturnian satellites]
{{\bf Properties of the small Saturnian satellites}: Mean radius $R_{\text{m}}$ in km from \citet{Thomas2013}; mass M in units of 10$^{19}$g taken from the JPL planetary satellite physical parameters file except for Janus, Epimetheus, Atlas, Prometheus, and Pandora from \citet{Jacobson2008}, Pan and Daphnis from \citet{Porco2007}, and Hyperion from \citet{Thomas2007}; mean density $\rho$ in kg/m$^3$ from Thomas et al. (2013; densities of Calypso, Telesto, and Helene are assumed and those of Aegaeon, Methone, and Pallene are inferred) except for Anthe and Polydeuces which are from the JPL planetary satellite physical parameters file\tablefootnote{https://ssd.jpl.nasa.gov/?sat\_phys\_par}; semi-major axis $a$ in km from \citet{Porco2007} except for Hyperion from \citet{Thomas2007} and Aegaeon from the JPL ephemeris file SAT342; surface gravity g in cm/$s^2$ and orbital velocity $V_s$ in km/s.}
\label{propsat}
\resizebox{\columnwidth}{!}{%
        \begin{tabular}{lrrrrrrrrr}
                
                Satellite & $R_{\text{m}}$ & M & $\rho$ & $a$ & g & $V_s$ \\ 
                \hline \rule{0pt}{2ex}Pan & 14.0 & 0.495 & 430 & 133584 & 0.17 & 16.85 \\
                Daphnis & 3.8 & 7.7x10$^{-3}$ & 340 & 136504 & 3.6x10$^{-2}$ & 16.67\\ 
                Atlas & 15.1 & 0.66 & 460 & 137670 & 0.19 & 16.59 \\
                Prometheus & 43.1 & 15.95 & 470 & 139380 & 0.57 & 16.49 \\
                Pandora & 40.6 & 13.71 & 490 & 141720 &  0.56 & 16.36 \\
                Epimetheus & 58.2 & 52.66 & 640 & 151410 & 1.04 & 15.83 \\
                Janus & 89.2 & 189.75 & 630 & 151460 & 1.57 & 15.83 \\
                Aegaeon & 0.33 & 6x10$^{-6}$ & 540 & 167425 & 4.9x10$^{-3}$& 15.05 \\
                Methone & 1.45 & 9x10$^{-4}$ & 310 & 194440 & 1.2x10$^{-2}$& 13.97 \\
                Anthe & 0.5 & 1.5x10$^{-4}$ & 500 & 197700 & 7x10$^{-3}$& 13.88 \\
                Pallene & 2.23 & 3x10$^{-3}$ & 250 & 212280 & 1.5x10$^{-2}$ & 13.37 \\
                Telesto & 12.4 & 0.674 & 500 & 294710 & 0.17& 11.35 \\
                Calypso & 9.6 & 0.315 & 500 & 294710 & 0.13& 11.35 \\
                Polydeuces & 1.3 & 4.5x10$^{-4}$ & 500 & 377200 & 1.8x10$^{-2}$ & 10.03 \\
                Helene & 18.0 & 1.139 & 400 & 377420 & 0.2 & 10.03 \\
        Hyperion & 135.0 & 561.99 & 544 & 1500933 & 2.05& 5.03 \\
\end{tabular}
}
\end{table}     
\end{center}    
\end{flushleft}

\vspace{-1cm}
\subsection*{A-ring satellites}
\begin{figure}[h!]
        \centerline{\scalebox{0.317}
                {\includegraphics{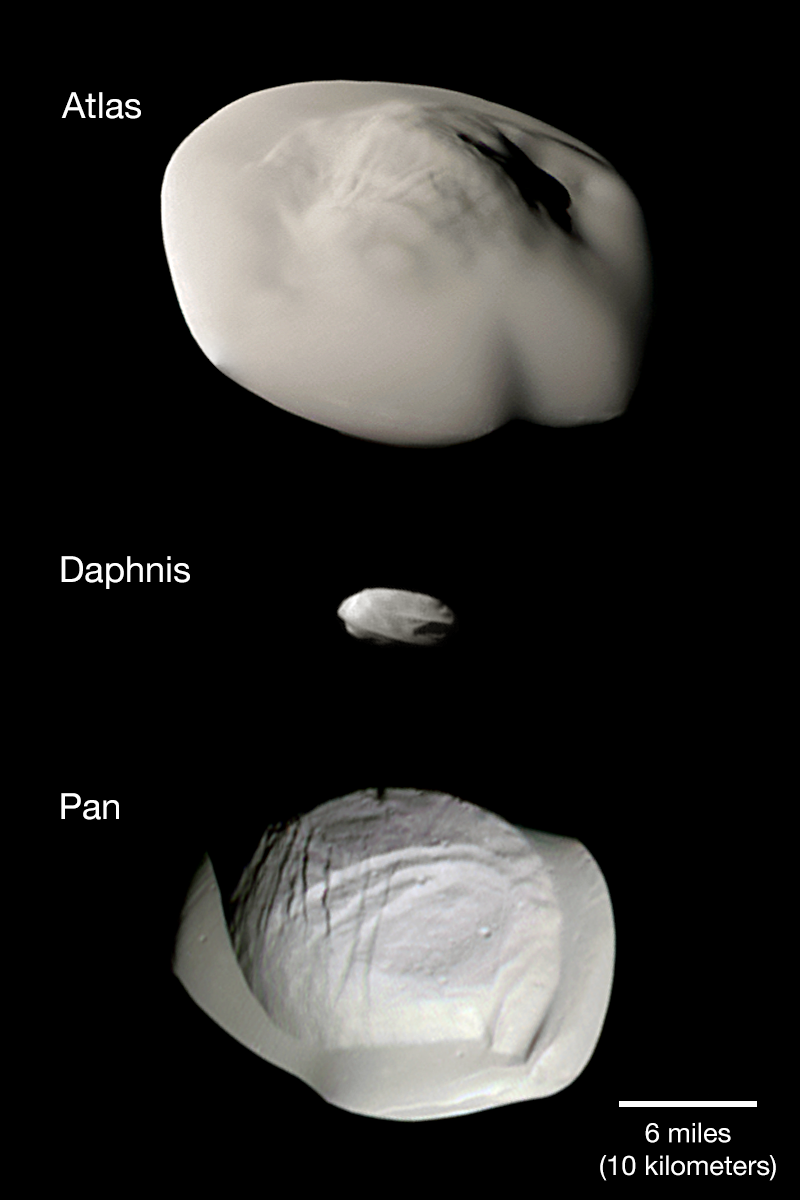}}}
        \caption{Montage of views from NASA's Cassini spacecraft of Pan, Atlas, and Daphnis. The images were taken using the Cassini spacecraft narrow-angle camera. Image Credit: PIA21449, NASA/JPL-Caltech/Space Science Institute.}
        \label{PanPIA}
\end{figure}

Pan ($R_{\text{m}}$ = 14 km) and Daphnis ($R_{\text{m}}$ = 3.8 km) orbit inside the A ring, the former within the Encke gap and the latter within the Keeler gap. Atlas ($R_{\text{m}}$ = 15.1 km) orbits just outside the A ring. These three satellites rotate at least approximately synchronously with their orbital periods and exhibit elongated shapes \citep{Thomas2013}. 

Observations for Pan and Atlas show that these objects have a distinct equatorial ridge, which  may be present on Daphnis as well \citep{Charnoz2007,Porco2007}. These features can be seen clearly in the striking images obtained by the Cassini mission 
(Fig. \ref{PanPIA}). The origin of these formations is unclear. On one hand, the rotational periods of these satellites are much too long (approximately 14 hours) for centrifugal force to compensate the gravitational force. On the other hand, tidal force generated by Saturn would deform bodies in the radial direction generating an ellipsoidal object and not a "flying saucer" object \citep{Charnoz2007}. According to \citet{Porco2007} one possible explanation is that all three satellites likely grew from cores that were one third to one half their present sizes by accumulation of A-ring material during an initial formation stage that took place inside a more vertically extended ring. Once they had filled their Roche lobe, a secondary stage of accretion formed their equatorial ridges in a disk that was already 20 meters thick (as in the current rings), which would explain the accumulation of particles along the equator. More recently, \citet{Leleu2018} found an alternative explanation considering head-on merging collisions between similar-sized bodies (also known as the pyramidal regime) as they migrated away from the rings. 

\subsection*{F-ring related satellites}
Prometheus ($R_{\text{m}}$ = 43.1km) and Pandora ($R_{\text{m}}$ = 40.6 km) orbit inside and outside the F ring, respectively. Originally they were both thought to be ring shepherds, but \citet{Cuzzi2014} proposed that the F ring is actually confined by Prometheus and precession effects. 

Both Prometheus and Pandora have distinctive elongated shapes. Prometheus presents a heavily cratered surface similar to those of Janus or Epimetheus and its topography suggests that it might be a partially delaminated object, with a region that could possibly be an exposed inner
core \citep{Thomas2013}. Pandora presents shallow craters, some apparently partly filled by ejecta and shallow craters similar to the ones present on Hyperion \citep{Thomas2013}. Additionally, Pandora has notable grooves up to 30 km in length and 1 km in width in the trailing part of the northern hemisphere, possibly due to tidal stresses and forced librations \citep{Morrison2009}. \par

\subsection*{Co-orbital satellites}

Among the satellites studied in this work, Janus ($R_{\text{m}}$ = 89.2 km) and Epimetheus ($R_{\text{m}}$ = 58.2 km) are the largest bodies after Hyperion, and exhibit a dynamic relation that is unique in our solar
system: they exchange orbits every 4 years due to their libration in horseshoe orbits \citep{Dermott1981,Yoder1989,Noyelles2010}, and at any given instant their semi-major axes differ by $\sim$ 48 km \citep{ElMoutamid2016}. These satellites may have originated after the breakup of a larger body during the early stages of the late heavy bombardment (LHB) \citep{Smith1982}. Their surfaces are heavily cratered showing craters in degradation states covered by loose material, but with distinct raised rims \citep{Thomas2013,Morrison2009}.

Epimetheus has grooves 5-20 km in length with varying widths in the south polar region \citep{Morrison2009}. Some of these grooves have straight, nearly parallel walls with well-defined breaks in slope between walls and floors which are characteristic of graben topography. The pattern of grooves is consistent with tensile stresses along the length of the object, most easily associated with tidal effects \citep{Thomas2013,Morrison2009}.
Janus was partially viewed at high resolution by Cassini, but it seems to have at least two  possible examples of single grooves as seen in the highest-resolution images \citep{Morrison2009}.

\subsection*{Embedded satellites}
Between the orbits of Janus/Epimetheus and Enceladus, four small satellites lie embedded in rings or arcs of debris.  Aegaeon ($R_{\text{m}}$ = 330 m), Methone ($R_{\text{m}}$ = 1.45 km), and Anthe ($R_{\text{m}}$ = 500 m) are all confined in their arcs by first-order MMRs with Mimas \citep{Hedman2010}, while Pallene ($R_{\text{m}}$ = 2.23 km) may librate about a third-order resonance with Enceladus \citep{Spitale2006} which allows ejecta material to spread freely and form a complete ring. 

Due to their recent discovery, there is yet no consensus on the origin of these satellites, but  according to several studies \citep{Hedman2009,Hedman2010, Sun2017} they are the most likely sources of their respective arcs or rings through the impact-ejecta process. Additionally, they all have smooth surfaces with apparently no presence of craters.

Aegaeon orbits within an arc near the inner edge of Saturn's G ring, trapped by the 7:6 corotation eccentricity resonance with Mimas \citep{Hedman2010}. Its surface is red and its geometric albedo is lower than 0.15, much darker than any other satellite interior to the orbit of Titan \citep{Hedman2011}. Considering that Aegaeon's associated arc is denser than the arcs related to Anthe or Methone, a recent impact may have caused Aegaeon to shed a significant amount of material, consequently leaving its darker interior exposed. Alternatively, the arc may contain debris with a broad range of sizes, perhaps the remains of a shattered moon \citep{Hedman2010,Hedman2011}.

Methone presents distinct albedo features on its leading hemisphere, which apparently are not correlated with variations in surface composition but instead could be indicators of variations in regolith grain size, soil compaction, or particle microstructure \citep{Thomas2013}. Additionally, Methone appears to sustain some process of fluidizing the regolith on geologic timescales to smooth impact craters, which results in a surface with no observed craters larger than 130 m \citep{Thomas2013}. 

\subsection*{Trojan satellites}
\begin{figure}[h!]
        \centerline{\scalebox{0.245}
                {\includegraphics{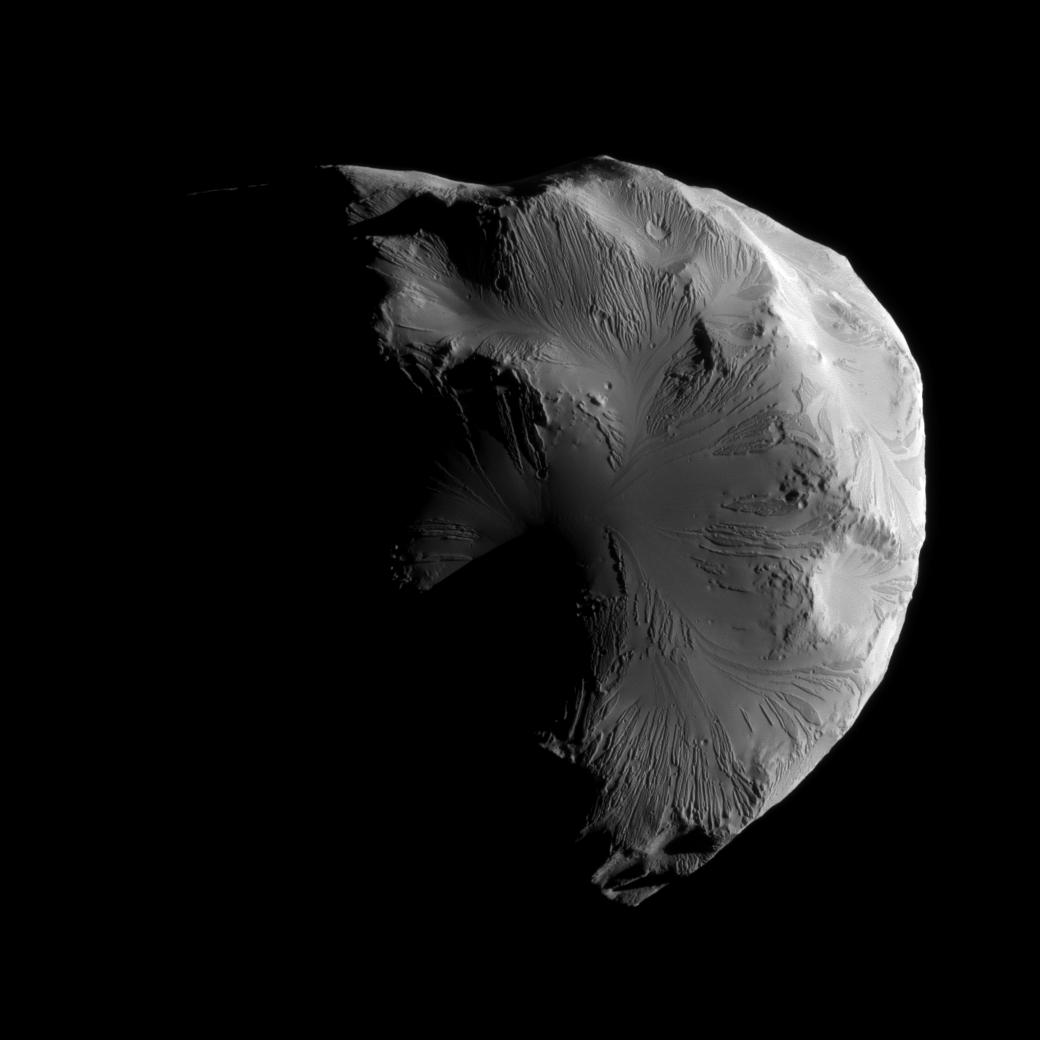}}}
        \caption{Leading hemisphere of Helene. The image scale is 42 meters per pixel and was taken in visible light with Cassini's narrow-angle camera. Image Credit: PIA12773, NASA/JPL-Caltech/Space Science Institute.}
        \label{HelenePIA}
\end{figure}

Among the mid-size Saturnian satellites, both Tethys and Dione each share their orbits with two trojan satellites. Telesto ($R_{\text{m}}$ = 12.4 km) and Calypso ($R_{\text{m}}$ = 9.6 km) librate around the leading and trailing lagrangian points of Tethys, respectively. Similarly, Helene ($R_{\text{m}}$ = 18km) and Polydeuces ($R_{\text{m}}$ = 1.3km) librate around the leading and trailing
lagrangian points of Dione, respectively.

Polydeuces is not well imaged, but according to high-resolution observations of Telesto, Calypso, and Helene, they all rotate synchronously with their orbital periods \citep{Thomas2013}.
These trojan satellites are thought to have formed via the process of mass accretion in an intermediate stage of the formation of Saturn's satellite system, where Tethys and Dione were almost formed, and the disk was already depleted of gas but had plenty of small planetesimals \citep{Izidoro2010}.

Telesto, Calypso, and Helene have large craters and evidence of covering of debris partially filling craters \citep{Thomas2013}. Even craters with diameters larger than 5 km appear to be buried \citep{Hirata2014}. Additionally, branching patterns of albedo and a topography that resembles drainage basins may indicate an undergoing downslope transport \citep{Morrison2009, Thomas2013}. This feature is characteristic of Saturn's trojan satellites and has not been observed on other small Saturnian satellites \citep{Thomas2013}.

As for the color properties of these satellites, Calypso and Helene show distinct bright and dark markings, also scarcely present on Telesto. This morphological diversity seems to be related to color diversity, relating different geological processes to different colors. Additionally, a trend of bluer colors with increasing distance from the main rings is visible, partially due to the influence of E-ring particles \citep{Thomas2013}.

According to \cite{Hirata2014}, when comparing the crater density on the leading side of Helene to that on its trailing side, the latter is ten times greater. This constitutes evidence for a bimodal surface, which can be explained if one takes into account the fine particle deposition over the leading side and its consequent erosion of craters, especially the smallest ones (See Fig. \ref{HelenePIA}). Those particles are thought to originate in the E ring, whose main source of material is Enceladus \citep{Filacchione2013}. Accordingly, for orbits far beyond Enceladus, like that of Helene, the velocities of satellites are expected to be greater than that of the E-ring particles, which favors the accumulation of these fine particles on Helene's leading side \citep{Hirata2014}.
Considering the large crater density on the trailing side and the cratering chronology by \citet{Zahnle2003}, \cite{Hirata2014} estimate that the surface age of the heavily cratered terrain is $\sim 4$ Gyr (and at least $\sim$ 1 Gyr), while the age of the E-ring particles deposit on the leading side is at most several tens of millions of years. Moreover, it is thought that Enceladus has not been active at its current level over the age of the solar
system but instead has had episodic geological activity \citep{Showman2013}. Hence, according to \cite{Hirata2014}, there is a possibility that the accumulation of E-ring particles on the surface of Helene began several million years ago as a result of the initiation of cryovolcanism of Enceladus.  

\subsection*{Hyperion}
\begin{figure}[h!]
        \centerline{\scalebox{0.248}
                {\includegraphics{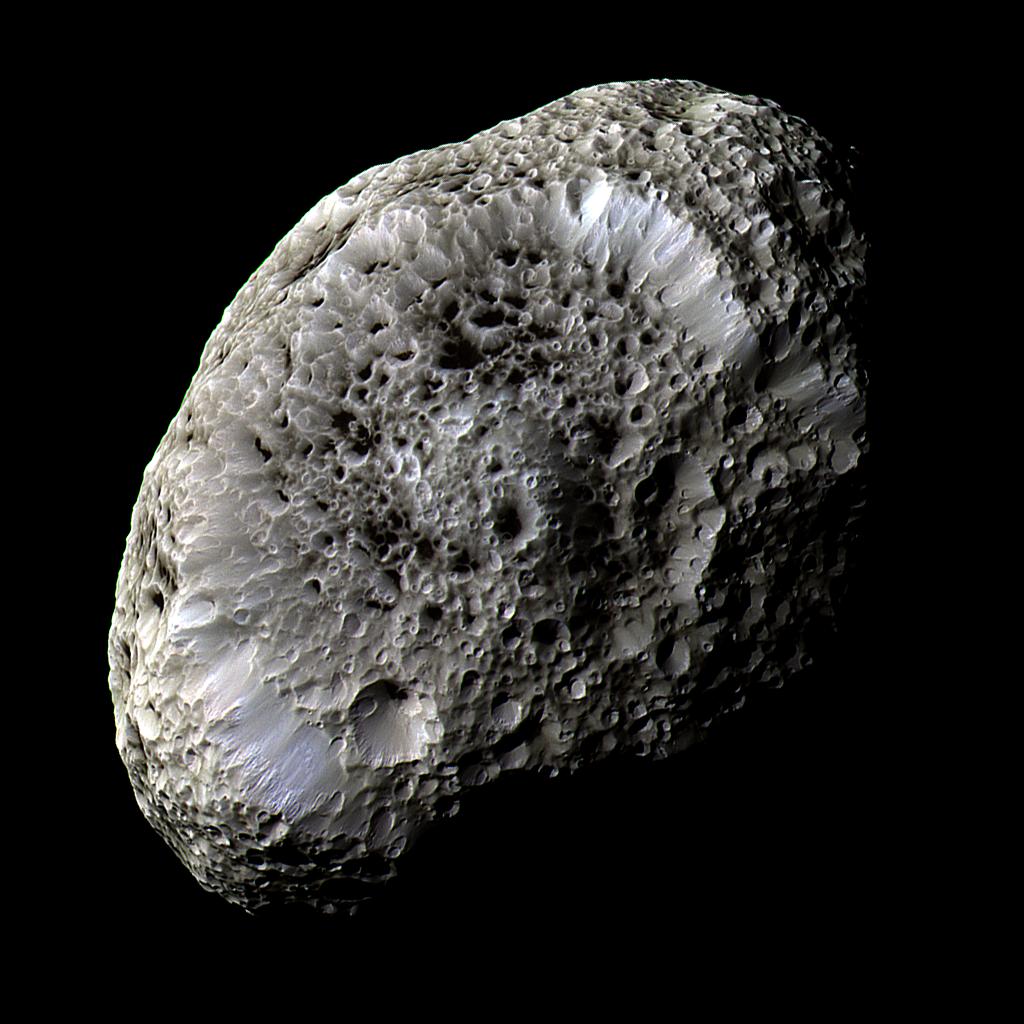}}}
        \caption{View of Hyperion produced by a combination of images taken using infrared, green, and ultraviolet spectral filters taken with Cassini's narrow-angle camera. The image scale is 362 meters per pixel. Image Credit: PIA07740, NASA/JPL/Space Science Institute.}
        \label{HyperionPIA}
\end{figure}
Hyperion  ($R_{\text{m}}$ = 135 km) is one of the most interesting and striking objects of the solar
system. It is the ``biggest'' of this group of small satellites and also the most ``distant'' from Saturn. It has an irregular shape (Fig. \ref{HyperionPIA}) and an unusually high degree of porosity ($>40 \%$; \citealt{Thomas2007}), estimated from its size and density and corroborated by the appearance of well-preserved craters 2-10 km in diameter (Fig. \ref{HyperionPIA}). \cite{Thomas2007} found that those craters, and also surviving short crater rims in heavily cratered zones of Hyperion, do not show erosion by superposed ejecta. 

Scaling laws and experiments on collisions on targets with porosities greater than $40 \%$ predict that the retained or produced ejecta are reduced by factors of more than four \citep{Housen2003}. Likewise, \citet{Howard2012} observed smooth surfaces and a reticulate, honeycomb pattern of narrow divides between craters, which they attribute to subsequent modification of crater morphology that occurs through mass-wasting processes accompanied by sublimation, probably facilitated by the loss of CO$_2$ as a component of the relief-supporting matrix of the bedrock. Both processes on Hyperion are thought to be the cause of its unique, ``sponge-like'' appearance. Cassini spectral observations of the surface of Hyperion in the ultraviolet and near-infrared reveal that its main component is a mixture of H$_{2}$O and CO$_2$ ice together with the presence of cyanide and complex organic material \citep{Cruikshank2007}.  

Cassini underwent four flybys in 2005 and 2006 at a distance of less than 300,000 km,  the closest one being at a distance of 618 km, from which it was possible to estimate the mass of Hyperion \citep{Thomas2007}. 

Chaotic rotation of Hyperion was inferred from Voyager observations \citep{Wisdom1984,Binzel1986} and was confirmed by Cassini, showing that the spin vector moves through the body and across the sky \citep{Thomas2007}. Its irregular form and chaotic rotation probably imply that Hyperion is a remnant of a larger body that was destroyed by a great impact in the past \citep{Smith1982}. 


\section{Method}
\label{method}
In this section, we present the method used to calculate the production of impact craters on the small satellites of Saturn. The impactor population, the collision calculations, and the cratering laws adopted for our model are set from our previous studies \citep{DiSisto2007, DiSisto2011, DiSisto2013}. We describe in detail the analytic method used to calculate the size of an impactor that catastrophically disrupted a satellite, the estimated age of the satellite, and, for those cases where observational crater counts are available, the surface age of the satellite.

\subsection{The crater production by Centaur objects}

\citet{DiSisto2011} and \citet{DiSisto2013} developed a theoretical model to calculate the cratering of a satellite due to impacts of Centaurs throughout the history of the solar
system considering its current configuration. 
The Centaur population considered in those works is the one obtained as a by-product of the dynamical evolution of SDOs through a numerical simulation done by \citet{DiSisto2007}. In that study, the authors considered a model of the SD and investigated the contribution of SDOs to the Centaur population. Therefore, we use the dynamical simulation of \citet{DiSisto2007} and the cumulative size distribution (CSD) of SDOs from the considerations of \citet{DiSisto2011}. In that paper, the authors analyzed the size-frequency distribution (SFD) of SDOs based on new estimations of the  maximum number of distant populations by \cite{Parker2010b,Parker2010a} and considered that it has a break at diameters $d \sim 60$ km  \citep{Bernstein2004,GilHutton2009, Fraser2009, Fuentes2008, Fuentes2009}. The CSD of our impactor population onto the small satellites of Saturn is therefore given by:
\begin{xalignat}{4}
        N(>d) &= C_0 \bigg(\frac{1 \text{km}}{d}\bigg)^{s_2 - 1} &&\text{for} && d \leq 60~\text{km},  \nonumber \\
        N(>d) &= \text{3.5} \times 10^{5} \bigg(\frac{100 \text{km}}{d}\bigg)^{s_1-1} &&\text{for} && d > 60~\text{km},
        \label{nr}
\end{xalignat} 
where $C_0 = \text{3.5} \times 10^{5} \times 100^{s_1-1} \times(60)^{s_2-s_1}$ by continuity for $d$ $=$ 60 km. 
As for the differential power law indexes, $s_1 =4.7$ for $d > 60$ km  \citep{Elliot2005} and given the uncertainty in the SFD at small sizes, two possible values are considered for $s_2$: $2.5$ and $3.5$ for $d < 60$ km. Therefore, we express our results in terms of both these exponents.
 
In order to evaluate the collisions of Centaurs on the small satellites of Saturn, we follow the method described in \citet{DiSisto2011} and \citet{DiSisto2013}. In those papers, the authors used the output files of the encounters of SDOs (those which are Centaurs) with Saturn, and, with a particle-in-a-box approximation, calculated the number of collisions between Centaurs and the satellites. Therefore, the cumulative number of collisions on the satellites depending on the impactor diameter is given by:
\begin{equation}
N_c(> d) =  \frac{v_i\,R_G^{2}}{v\,(R_{\text{H}})\,R_{\text{H}}^{2}}10.257\,N(> d),
\label{Nc10}
\end{equation}
where $v_i$ is the relative collision velocity on each satellite, v($R_{\text{H}}$) is the mean relative encounter velocity of the Centaurs when they enter the Hill sphere (of radius $R_H$) of Saturn, and $R_G$ is the satellite's gravitational radius of collision given by $R_G=R_s (1+(v_{esc}/v(R_{\text{H}}))^2)^{1/2}$. 
We have considered here $R_G$ instead of the physical radius $R_s$ as in the previous papers, although
since the encounter velocities are high, $R_G$ is only slightly greater than $R_s$.
The relative encounter velocity was calculated from the encounter files obtained in \citet{DiSisto2007}. The relative collision velocity on each satellite $v_i$ was calculated by assuming that collisions on the satellite are isotropic, and then $v_i = \sqrt{v_s^2+v_0^2}$, where $v_s$ is the velocity of the satellite and $v_0$ is the mean relative velocity of Centaurs when they cross the orbit of a satellite. 
With Eq. \ref{Nc10}, it is possible to obtain the number of collisions on a given satellite in terms of the size of the impactor. Subsequently, in order to relate the number of collisions with the size of the crater produced on the satellite, we use the scale relation between the projectile size and the crater size from \citet{Holsapple2007}. In this work, the authors presented the updated scaling laws for cratering in a work focused on interpreting the observations of the Deep Impact event. The diameter $D_{\text{t}}$ of a crater produced by an impactor of diameter $d$ can be obtained from the general equation \citep{Holsapple2007}:
\begin{equation}
D_{\text{t}} = K_{1}\left[\left(\frac{gd}{2v^{2}_{\text{i}}}\right)\left(\frac{\rho_{\text{t}}}{\rho_{\text{i}}}\right)^{\frac{2\nu}{\mu}}
+ K_{2}\left(\frac{Y}{\rho_{\text{t}}v^{2}_{\text{i}}}\right)^{\frac{2+\mu}{2}}\left(\frac{\rho_{\text{t}}}{\rho_{\text{i}}}\right)^{\frac{\nu(2+\mu)}{\mu}}\right]^{-\frac{\mu}{2+\mu}} d,
\label{HH07}
\end{equation}
where the two exponents $\nu$ and $\mu$ and the constants $K_1$ and $K_2$ characterize different materials. 
The surfaces of the satellites of Saturn are mainly composed of H$_{2}$O ice. Therefore, we use the 
cratering law for icy surfaces for all satellites following the analysis made by \cite{DiSisto2013} from calculations of the scaling laws by \cite{Kraus2011} onto H$_{2}$O ice. For the strength parameters ($K_2$ and $Y$), we consider values for H$_{2}$O ice from \cite{DiSisto2013}. Thus, 
 the values of the parameters are $\mu=0.38$, $\nu=0.397$, $K_1 = 1.67$, and $K_2 = 0.351$ and for the strength $Y = 1.5 \times 10^5$ dyn/cm$^2$. 
From Eq. \ref{HH07}, it is possible to obtain the crater diameter ($D$) for a given impactor diameter ($d$). With Eq. \ref{Nc10} we obtain the cumulative number of collisions for this given impactor, and subsequently the cumulative number of craters for that crater diameter $N_c(> D)$.

The first term in Eq. \ref{HH07} is a measure of the gravity of the target and the second term indicates the importance of the strength of the target. Thus, if the first term is larger in value than the second term, the crater is under the gravity regime, whereas if the second term is larger, it is under the strength regime. The partition between the two size scales of impacts depends on the size of the event \citep{Holsapple1993}. When the two terms are equal we obtain the transition impactor diameter ($d_{\text{l}}$) between the strength regime and the gravity regime. From Eq. \ref{HH07} we can calculate the crater diameter ($D_{\text{l}}$) produced by $d_{\text{l}}$ , which is shown in Table \ref{rgrales}. Additionally, we consider the most probable impact angle $\theta = 45^{\circ}$, and therefore the impact velocity is multiplied by $sin(45^{\circ})$.

From   Eq. \ref {HH07},  we have  $D \propto d$  for craters with diameters $D < D_l$, that is in the strength regime. Thus, the cumulative crater size distribution ($N_c(>D)$) will follow the same power-law relation given for impactors (Eq. \ref {nr}).  In the gravity regime, where  $D > D_l$, the relation between $D$ and $d$ is no longer linear but $ D \propto d^{\frac{2}{2+\mu}}$ . Subsequently, the dependence of $N_c(>D)$ on $D$ considering both cratering regimes separately will be given by: 
\begin{xalignat}{4}
N_c(>D) & \propto D^{-s_2 + 1} &&\text{for} && D < D_l,  \nonumber      \\
N_c(>D) & \propto D^{-s_c}        &&\text{for} && D > D_l,        
\label{nd1}
\end{xalignat} 
where we have only considered the case for $s_2$ of the CSD because all impactors on the small satellites have diameters below 60 km, and $ s_c = (1+{\frac{\mu}{2}})(s_2 - 1)$. Therefore, if $s_2 = 2.5$, $s_c = 1.786$ and 
if $s_2 = 3.5$, $s_c = 2.976$.

Equation \ref{HH07} sets the transient radius of a crater for an impactor of diameter $d$. However, some degree of slumping and mass movement makes the final crater wider and shallower than the transient crater. Therefore, we consider the diameter of the final simple crater to be $D = 1.3 k D_t$, where $k = 1.19$ from \cite{DiSisto2013}.
However, above a certain threshold size (which depends on the target), a simple crater  collapses because of gravitational forces,  ultimately leading to {\it complex} craters with central peaks, terraced walls, and circular rings. We follow the treatment in \citet{Kraus2011} as in \citet{DiSisto2013} to obtain the final crater size. However, the crater collapse or deformation is strongly dependent on the target gravity and then the simple-to-complex transition diameter adjusted inversely with gravity: 
\begin{equation}
D^{*} = \frac{2g_{gan}}{g}, 
\end{equation}
where  $g_{gan} = 142.8 \,$ cm/s$^2$ is the surface gravity of Ganymede and $g$ the surface gravity of the satellite, which is shown in Table \ref{propsat}. 

Since the small satellites of Saturn have a very low surface gravity (due to their low mass), $D^*$ is high, and is in fact greater than the diameter of the satellite in all cases except for Hyperion. However, in this case the largest theoretical crater is also smaller than $D^*$. Therefore, all craters produced on the small satellites are simple craters. 

\subsection{Disruption of satellites}
\label{d}
Considering that a number of the studied satellites here are relatively small in size (e.g., Aegaeon, Anthe), it is convenient to study those cases in which a catastrophic collision could lead to their disruption. In order to determine this, we have based our calculations on the work of \cite{Benz1999}, where a hydrodynamic particle method is used to simulate collisions between rocky and icy bodies of several sizes. Their goal is to find a threshold for the catastrophic disruption.   
Impacts can be grouped in different categories depending on their outcome, which are also dependent on the regime (strength or gravity) under which the collision occurs. The experiments of \cite{Benz1999} included both strength and gravity regimes, however they found that gravity plays a dominant role on the collision result, even on small targets. Therefore, we consider \textit{dispersing} collisions, that is events that break the parent body into smaller pieces, and manage to impart velocities to those fragments in excess of escape velocity. 

Therefore, we adopt the specific energy threshold used in the literature as the kinetic energy in the collision per unit mass for a dispersing event $Q^*_D$, which is the specific energy required to disperse the targets into a spectrum of individual but possibly reaccumulated objects, the largest one having exactly half the mass of the original target.
Based on laboratory experiments for different sizes and collision velocities, \cite{Benz1999} fit an analytical curve for $Q^*_D$ of the functional form:
\begin{equation}
Q^*_D=Q_{\text{0}}\left(\frac{R_{\text{pb}}}{1{\text{cm}}}\right)^{a}+B\,\rho\left(\frac{R_{\text{pb}}}{1{\text{cm}}}\right)^{b}
\label{Rup3}
,\end{equation}
where $R_{\text{pb}}$ is the radius of the parent body (or target) in centimeters (cm), $\rho$ the density of the parent body in g/cm$^3$, 
and $Q_{\text{0}}$, $B$, $a$ and $b$ are constants that the authors determine for various materials. We adopt the values for ice and impact velocities of 3 km/s since these are the most similar material and velocity studied. Therefore $Q_{\text{0}}$=1.6x10$^7$erg/g, $B$=1.2 erg.cm$^3$/g$^2$, $a$=-0.39 y $b$=1.26.

The kinetic energy in the collision per unit mass is given by:
\begin{equation}
Q=\frac{1}{2}\bigg(\frac{M_{\text{p}}}{M_{\text{pb}}+M_{\text{p}}}\bigg)\,v^{2}_{\text{i}}
\label{Rupa}
.\end{equation}

Considering that the largest remnant has a mass equal to half the mass of the original target, then $Q = Q^*_D$. Therefore, from  Eqs. \ref{Rup3} and \ref{Rupa}, the mass of the projectile that catastrophically disrupts a target is given by:

\begin{equation}
        M_{\text{p}}=\bigg(\frac{2Q^*_DM_{\text{pb}}}{v^{2}_{\text{i}}-2Q^*_D}\bigg)
        \label{Rup6}
        ,\end{equation}
which for a spherical impactor gives the impactor radius that generates the target disruption:
\begin{equation}
R_{\text{p}}=\left(\frac{6Q^*_DM_{\text{pb}}}{4\pi 
        \rho_{\text{p}}(v^{2}_{\text{i}}-2Q^*_D)}\right)^{\frac{1}{3}}
\label{Rup77}
.\end{equation}

\subsection{Satellite age}
\label{age}
After calculating the number of impacts suffered by each of the studied satellites and analyzing which of those impacts leads to their disruption, we find that for some satellites more than one catastrophic collision takes place over the age of the solar
system (see Sect. \ref{result}). Therefore, these satellites must be younger than the age of the solar
system. As a first-order assumption, we consider that they must have formed or re-formed after the last catastrophic collision. However, there are other possibilities that we discuss in detail for the corresponding cases in Sect. \ref{result}.  

After finding the threshold for the catastrophic disruption of a given satellite, we find it natural to calculate the time when the last catastrophic collision took place. This would be the ``age'' of the satellite.

The cumulative number of craters produced by Centaurs on the satellites is proportional to the number of encounters between Centaurs and Saturn \citep{DiSisto2011} (see Eq. \ref{Nc10}). Therefore, the dependence of cratering with time is the same as the dependence of encounters with time. \citet{DiSisto2016} found that this time dependence is well fitted by a logarithmic function given by:

\begin{equation}
F(t)=a\ln{t}+b
\label{edad1}
,\end{equation}
where $a$=0.198406 $\pm$ 0.0002257 and $b$=-3.41872 $\pm$ 0.004477.

Thus, the cumulative number of craters for a given satellite as a function of time can be obtained from the following equation.
 \begin{equation}
          N_c(>D,t) = F(t)N_c(>D) 
          \label{edad2}
,\end{equation}
where $N_c(>D)$ is the total number of craters that have a diameter larger than $D$ produced over the age of the solar
system (Eqs.~\ref{Nc10} and \ref{HH07}).
Thus, the last collision that catastrophically fragmented the satellite occurred at a time $t$ that can be obtained from the following equation.

\begin{equation}
N_c(>D_{rup},t_f) - N_c(>D_{rup},t)=1
\label{edad3}
,\end{equation}    
where $t_f=$ 4.5 Gyr is the age of the solar
system and $D_{rup}$ is the crater diameter that would correspond to a catastrophic collision.
Combining the expressions \ref{edad1} and \ref{edad2} with \ref{edad3} we obtain:
\begin{equation}
N_c(>D_{rup})[a(\ln{t_{\text{f}}}-\ln{t})]=1
.\end{equation}
The age of the satellite corresponds to the time span from its formation to the present. Therefore, the age ($\tau$) of the satellite is given by: 

\begin{equation}
\tau=t_f-t=t_f\Big(1-e^{-\frac{1}{aN_c(>D_{rup})}}\Big) 
\label{edad11}
.\end{equation}
Since our model predicts the cratering of satellites over the total age of the solar
system, for those cases where disruption occurred it is convenient to adapt the cratering model to the age of the corresponding satellite. 
 
Therefore, we consider Eqs. \ref{edad1} and \ref{edad2} and the ages $\tau$ calculated for the satellites we studied.
For a satellite of age $\tau $ which fragmented at time $t$, the cumulative number of craters produced in the last $t_f - t$ years is therefore given by: 
\begin{equation}
N_c(>D) - N_c(>D,t) = [1 - F(t)]N_c(>D)
\label{nuevaprod}
.\end{equation}

\subsection{Surface ages or cratering timescale}
\label{surfaceage}

From the method described in the previous subsections, we can calculate the theoretical number of craters on a given satellite ($N_c(>D)$) produced throughout the age of the solar
system. For those satellites that have crater counts available (from Voyager or Cassini observations), a comparison with our results is possible. Consequently, we test the cratering model but also infer if there may be physical processes that affect the satellite as a whole or its surface features. In this regard, \cite{DiSisto2016} analyzed the cratering process in the mid-sized saturnian satellites and calculated their surface ages. The idea is very simple and should be considered a first-order approach to the problem. If $N_0(>D)$ is the cumulative number of observed craters on a given satellite, then if $N_0(>D) > N_c(>D)$ for all $D$, the satellite may be primordial and has preserved all the craters over its 4.5 Gyr age. On the contrary, if $N_0(>D) < N_c(>D)$, at first one can say that there may have been geological or endogenous processes that erased some craters. But, if $N_0(>D) < N_c(>D)$ for all values of $D$, and one cannot detect any erosive processes, the satellite may be young, meaning it could have formed in recent times.  

To consider this matter, we follow the method developed in \cite{DiSisto2016} to calculate the age of a satellite surface. Therefore the estimated cratering timescale or the ``age'' of the surface $\tau_s$ is given by:
\begin{equation}
\tau_s(>D) = t_f (1-e^{-\frac{N_o(>D)}{a N_c(>D)}} ) , 
\label{agesup} 
\end{equation}
where $t_f = 4.5$ Gyr is the age of the solar
system. 

We note that this formula is similar to Eq. (\ref{edad11}) because it was obtained considering Eq. (\ref{edad3}), but scaled to the observed number depending on $D$, therefore:   
\begin{equation}
N_c(>D,t_f) - N_c(>D,t)= N_0(>D)
\label{n}
.\end{equation}    
Hence, $\tau_s(>D)$ represents the time span from the present, in which craters were produced on the satellite with the cratering rate of our model. Thus, if there are physical processes that erase craters, then $\tau_s(>D)$ is the age of the surface for each diameter $D$ and indicates the scale of time during which these physical processes acted for each $D$. On the contrary, if there are no physical processes (or they are negligible), and $N_0(>D) < N_c(>D)$ for all values of $D$, then the satellite must have formed $\tau_m$ years ago, where $\tau_m$ is the maximum value of $\tau_s(>D)$.  

We explore this subject for each particular case in the following section. 

\section{Results}
\label{result}

Following the method and formulations described in Sect. \ref{method}, we calculate the cratering production on the small satellites over the age of the solar
system by Centaur objects. Our results are shown for the two values of the impactor size distribution index ($s_2 = 2.5$ and $s_2 = 3.5$) and the cumulative number of craters per square kilometer is plotted for each satellite. The cumulative number of craters on the whole surface of the satellite that  we obtained from our model (Eqs. \ref{Nc10} and \ref{HH07}) was divided by the surface area of a sphere of equal radius to the mean radius of the satellite (see table \ref{propsat}). The lower limit of the plots corresponds to the largest crater produced on the whole surface of the satellite. The values of the largest impactor diameter ($d_m$) and the largest crater diameter generated by that impactor ($D_m$), the transition crater diameter between strength and gravity regimes ($D_l$), and the collision velocity on each satellite were calculated and are shown in Table \ref{rgrales}.  As mentioned in Sect. \ref{method}, some of these satellites are very small; this is the case for Pan, Daphnis, Atlas, Aegaeon, Methone, Anthe, Pallene, Calypso, and Polydeuces. We find that these satellites suffer one or more catastrophic collisions during the age of the solar
system for $s_2 = 3.5$. Consequently, we calculate the diameter of the impactor that catastrophically disrupted them and list their physical diameters for reference in Table $\ref{rgrales2}$. Further, according to our calculations these satellites may be younger than the solar
system age. As a first-order assumption, we consider that they must have formed (or re-formed) after the last catastrophic collision. Therefore their ages were calculated by Eq. \ref{edad11} and are shown in Table \ref{rgrales2}. In these specific cases we also calculate the cumulative number of craters produced during the real age of the satellite using Eq.\ref{nuevaprod}.

In order to analyze and compare our results, we use the crater counts available in the literature, mainly based on Cassini observations. Most of the crater counts used in this work were presented in \citet{Thomas2013} (provided by Peter C. Thomas, personal communication, October, 2017), where the authors analyzed the physical and dynamical characteristics of the small Saturnian satellites. For Telesto and Helene we also used crater counts plotted in \citet{Hirata2014} (provided by Naoyuki Hirata, personal communication, November, 2017), where the authors studied the bimodality of the surface of Helene and its correlation with the E-ring particles. Crater counts on the surface of  Hyperion were first presented in \citet{Thomas2007} from Cassini observations and were provided by Peter C. Thomas, personal communication, October, 2017. Additionally, we included crater counts presented in \citet{Plescia1985}, where the authors modeled the impact flux history of the Saturnian system based on observed crater density data from the Voyager 1 and 2 missions.

In general, crater counts were done for specific areas in each satellite, and our model considers the cratering process over the whole surface. Thus, the comparison between our model and the available observations can only provide information on the peculiarities of that specific zone.

In the following subsections we analyze our results for each satellite, considering that their characteristics and environments are different in each case. 

\begin{table}[h]
\caption[General results]
{{\bf General results}: Collision velocity $v_i$ in kilometers per second, transition crater diameter between the strength and gravity regimes $D_l$, largest impactor diameter $d_m$ and largest crater diameter $D_m$ on the  surface of the satellite due to the Centaur population, considering both indexes of the distribution $s_2$=2.5 and $s_2$=3.5, all expressed in kilometers. For Aegaeon, the absence of impact craters for the distribution index $s_{\text{2}}$=2.5 is indicated by a dash `-'.}
\label{rgrales}
\begin{flushleft}
        \resizebox{\columnwidth}{!}{%
        \begin{tabular}{lrrrrrr}
                Satellite & $v_i$ & $D_l$  & $d_m$ & $D_m$ & $d_m$ & $D_m$ \\
                &   &    & $s_2$=2.5 & $s_2$=2.5 & $s_2$=3.5 & $s_2$=3.5 \\ \hline
                \rule{0pt}{2.6ex}Pan & 29.48 & 25.67 & 0.02 & 2.23 & 0.55 & 45.91 \\ 
                Daphnis & 29.17 & 153.78 & 4x10$^{-3}$ & 0.4 & 0.19 & 18.83 \\ 
                Atlas & 29.04 & 20.94 & 0.02 & 2.4 & 0.58 & 46.5 \\
                Prometheus & 28.87 & 6.91 & 0.1 & 8.85 & 1.33 & 83.72 \\
                Pandora & 28.63 & 6.84 & 0.1 & 8.08 & 1.26 & 79.23 \\
                Epimetheus & 27.73 & 2.80 & 0.15 & 10.6 & 1.66 & 82.6 \\
                Janus & 27.73 & 1.85 & 0.27 & 16.55 & 2.34 & 103.69 \\
                Aegaeon & 26.4 & 938.11 & - & - & 0.03 & 2.26 \\
                Methone & 24.55 & 210.33 & 1x10$^{-3}$ & 0.09 & 0.08 & 7.82 \\
                Anthe & 24.4 & 93.03 & 2x10$^{-4}$ & 0.02 & 0.03 & 3.01 \\
                Pallene & 23.52 & 168.29 & 2x10$^{-3}$ & 0.16 & 0.11 & 11.13 \\
                Telesto & 20.08 & 12.63 & 0.01 & 1.23 & 0.42 & 29.16 \\
                Calypso & 20.08 & 16.31 & 0.01 & 0.87 & 0.35 & 24.75 \\
                Polydeuces & 17.86 & 209.64 & 6x10$^{-4}$ & 0.05 & 0.07 & 5.07 \\
                Helene & 17.85 & 10.88 & 0.02 & 1.87 & 0.55 & 35.99 \\
                Hyperion & 9.62 & 1.66 & 0.23 & 10.36 & 2.14 & 68.93 \\
        \end{tabular}
        }
\end{flushleft}
\end{table}
\begin{table}[h]
        \caption[Results for fragmented satellites]
        {{\bf Results for fragmented satellites}: Mean satellite diameter $D_s$ in kìloilometers, diameter of the last disruptive impactor $d_{rup}$ in kilometers, and estimated age of the satellite $\tau$ in gigayears.}
        \label{rgrales2}
        \begin{flushleft}
                \begin{tabular}{lrrrr}
                Satellite & $D_s$ &$d_{rup}$ & $\tau$ \\ \hline
                \rule{0pt}{2.6ex}Pan & 28.0 & 0.4 & 4.04 \\ 
                Daphnis & 7.6& 0.06 & 0.9 \\ 
                Atlas & 30.2& 0.47 & 4.27 \\
                Aegaeon & 0.7& 3x10$^{-3}$ & 0.12 \\
                Methone & 2.9& 0.01 & 0.32 \\
                Anthe & 1.0 & 5x10$^{-3}$ & 0.19 \\
                Pallene &4.5&  0.02 & 0.43 \\
                Calypso & 19.2& 0.33 & 4.45 \\
                Polydeuces & 2.6& 0.02 & 1.03 \\    
                \end{tabular}
        \end{flushleft}
\end{table}
\subsection{A ring satellites}
\begin{figure}[!h]
        \centerline{\scalebox{0.7}
                {\includegraphics{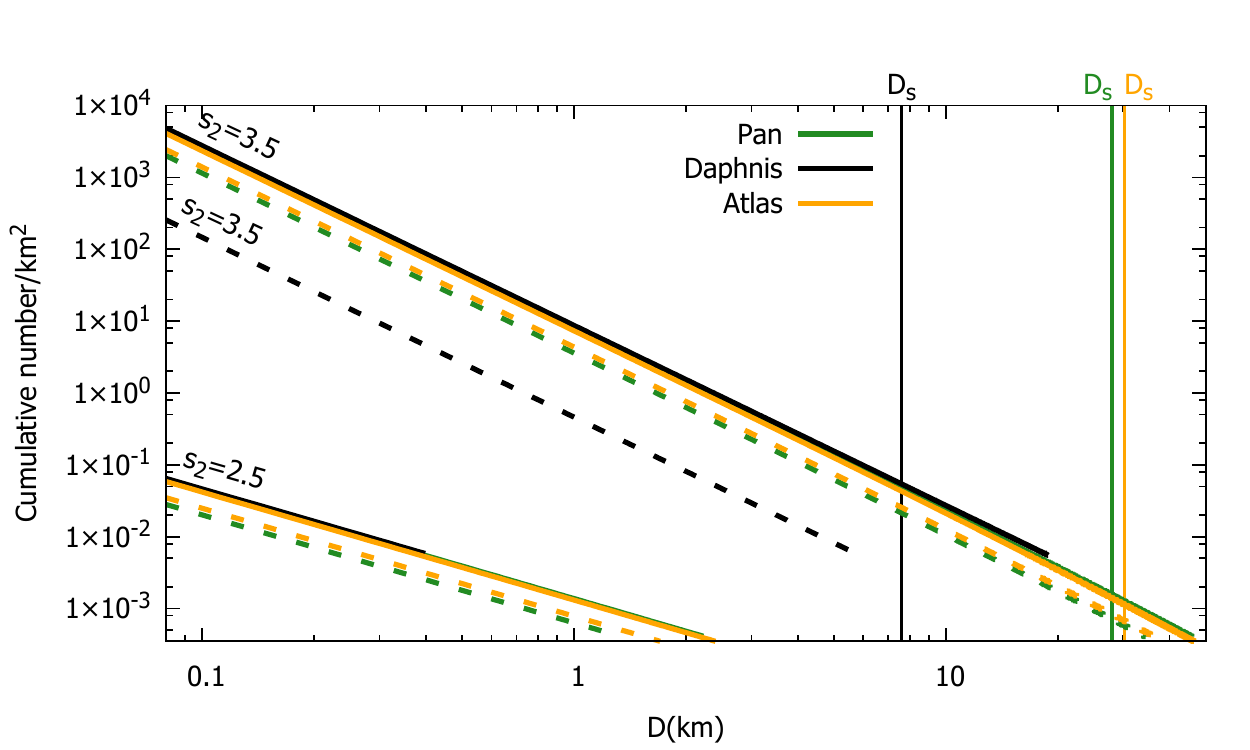}}}
        \caption{Cumulative number of craters per square kilometer as a function of crater diameter $D$ for Pan, Daphnis, and Atlas, calculated for both values of the impactor size distribution index $s_2$. Solid lines correspond to the cratering process considering the total age of the solar
system and dashed lines correspond to the adapted model for the calculated age of the satellite (top lines are for $s_2$=3.5 and bottom lines for $s_2$=2.5). Vertical lines indicate the mean body diameter $D_s$ for each satellite.}
        \label{SatA}
\end{figure}
The cumulative number of craters per square kilometer for the A-ring satellites is shown in Fig. \ref{SatA}.
Our results for Pan, Atlas, and Daphnis show that all three satellites suffer catastrophic collisions. Consequently, we estimate their ages to be $\sim$4 Gyr for Pan and Atlas and $\sim$0.9 Gyr for Daphnis.

Our adapted cratering model considering the age of each of these satellites does not display major changes in Pan and Atlas, given that their estimated age is similar to the age of the solar
system. For the distribution index $s_2$=3.5, the largest crater in Pan is 34.5 km in diameter and in Atlas is 38.5 km in diameter. On the other hand, our calculations for Daphnis show that it might have suffered several
catastrophic collisions, reducing its estimated age to $\sim$0.9 Gyr. Thus, assuming that the satellite was formed 0.9 Gyr ago, the adapted cratering curve changes considerably (see Fig. \ref{SatA}). The largest crater in the adapted model is 5.85 km in diameter. Additionally, for the adapted model we find that all the craters in these three satellites were produced under the strength regime.

Although there are no crater counts available for these satellites, observations made by the Cassini mission show that their surfaces might experience strong modifications related to the proximity to the A ring particles, which could be modifying and eroding craters on relatively short time scales. 
Moreover, as mentioned previously,  the equatorial ridges of these satellites may have formed via A-ring particle accretion, and therefore the interaction between these satellites and the A ring may represent the main source of the physical processes on their surfaces.   

\subsection{F-ring-related satellites}

Prometheus and Pandora orbit in a chaotic way at both sides of the F ring, due to the superposition of various resonances. This causes Prometheus to have periodic encounters with the ring via gravitational interactions, consequently leaving a set of streamers. These interactions most probably leave traces on the surface of the  satellite, eroding its craters and adding to the sediment accumulation.

\begin{figure}[h!]
        \centerline{\scalebox{0.7}
                {\includegraphics{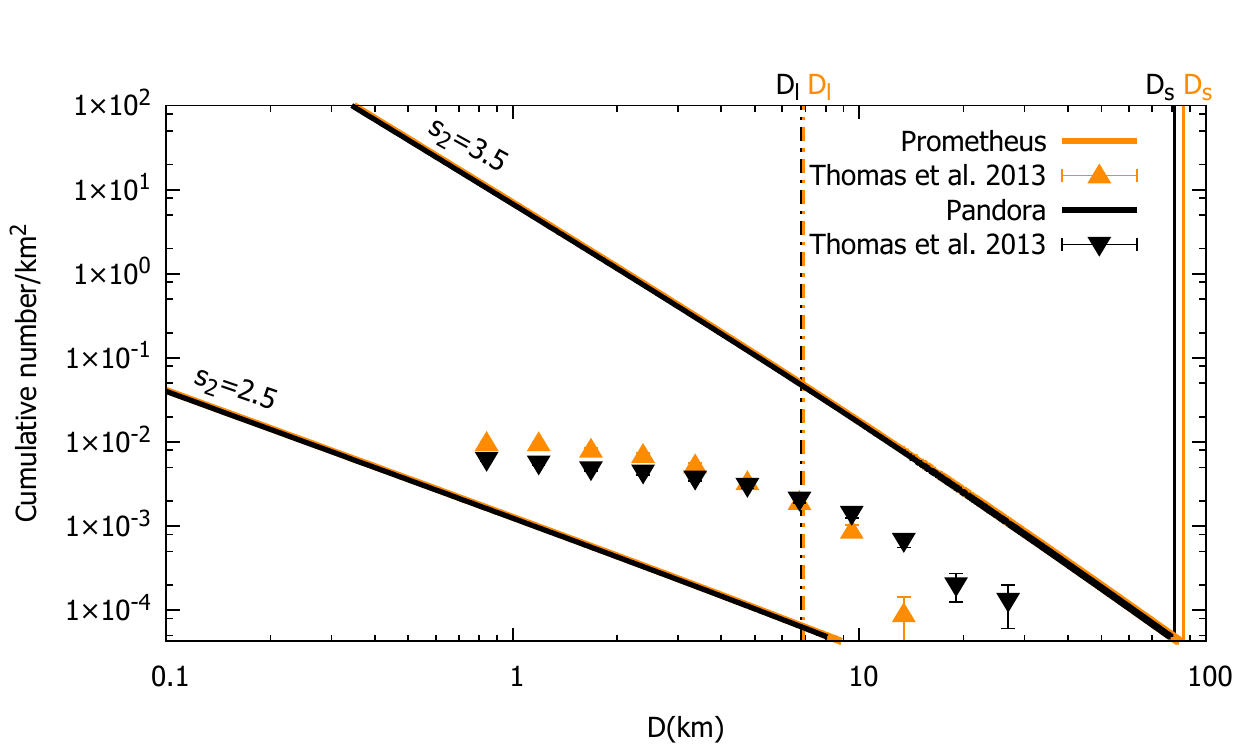}}}
        \caption{Cumulative number of craters per square kilometer as a function of crater diameter $D$ for Prometheus and Pandora. Solid lines refer to our model calculated for both values of the impactor size distribution index (top lines are for $s_2$=3.5 and bottom lines for $s_2$=2.5) and color points represent the observations presented in Thomas et al. (2013). Dashed vertical
lines indicate the strength-to-gravity transition crater diameter $D_l$ on each satellite and solid vertical
lines indicate the mean body diameter $D_s$ for each satellite.}
        \label{SatF}
\end{figure}

Figure \ref{SatF} shows the theoretical cumulative number of craters for the two indexes $s_2$ and the crater counts from Cassini observations presented in \citet{Thomas2013} that correspond to partial areas of Prometheus (imaged area of 11,400 km$^2$) and Pandora (imaged area of 14,900 km$^2$). The error bars presented in the plots are statistical errors proportional to 1/$\sqrt{n}$, where $n$ is the cumulative number of craters.

As can be seen in Fig. \ref{SatF}, observations lie in between both theoretical curves, but they are more similar to our model for $s_2$=3.5. Besides, our model for the $s_2$=2.5 index of the impactor population does not predict any of the observed craters. Therefore, we discard the results for the $s_2$=2.5 index and consider only $s_2$=3.5 for the comparison with the observations.
 
In this case, we find that the largest crater produced on these satellites is $\sim$80 km in diameter, while the largest observed craters are 26.9 km in diameter (Pandora) and 13.45 km in diameter (Prometheus). In addition, the whole theoretical curve is above the observed one. As mentioned, given the proximity to the F ring, we should consider F-ring particle deposition on the surface of these satellites, covering the craters and possibly erasing them, consequently causing the observed count to be lower. As for the smaller craters ($D \lesssim 8$ km), our model deviates more from the observed curve. A possible explanation for this is that erosion effects are larger on the smaller craters, but we should also consider crater saturation \citep{Richardson09}, due to the fact that large craters could erase any traces of smaller craters previously present in the surface. 

\begin{figure}[h!]
        \centerline{\scalebox{0.7}
                {\includegraphics{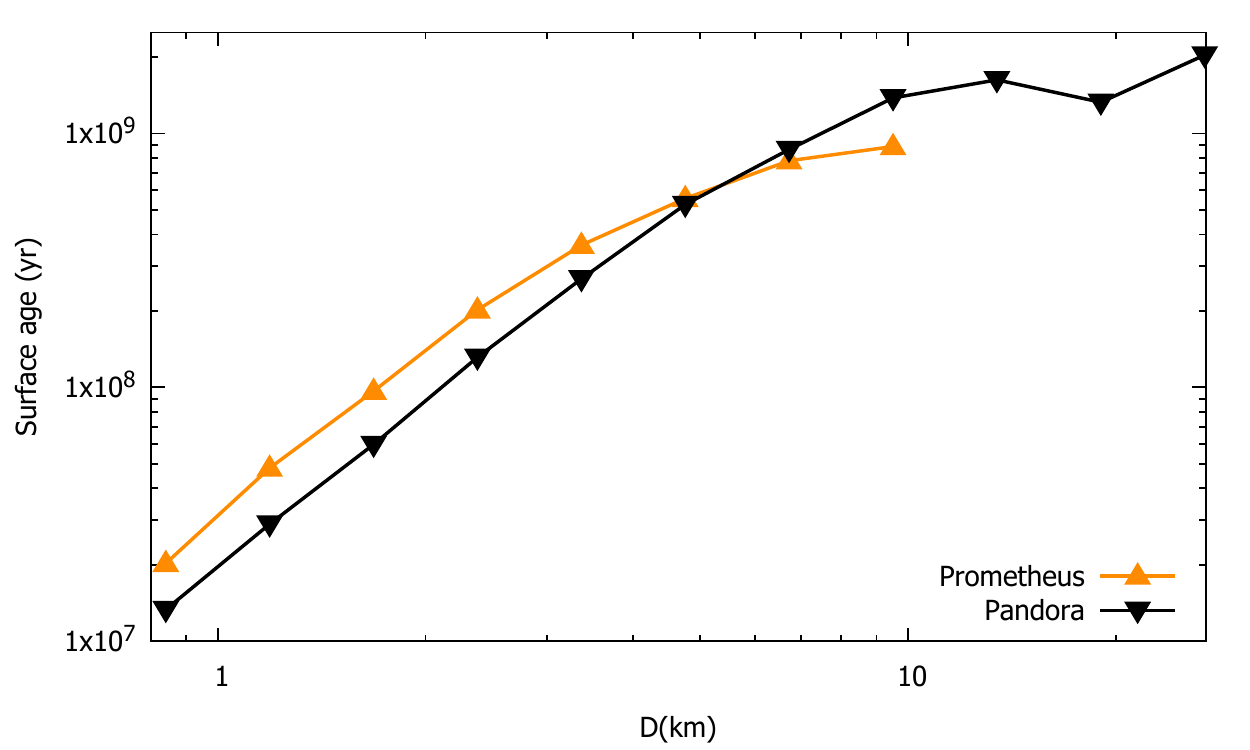}}}
        \caption{Surface age as a function of crater diameter $D$ for Prometheus and Pandora.}
        \label{edads_pp}
\end{figure}

Alternatively, given that the difference between observed and theoretical crater counts extends for the whole range of crater diameters, there is a possibility that these satellites are not primordial but instead formed recently. Regardless, the calculation of the surface age (Sect. \ref{surfaceage}) limits the time span during which the erosion process acted or the age of formation or capture on its current orbit, as discussed in Sect. \ref{surfaceage}. Surface ages for those satellites as a function of diameter can be seen in Fig. \ref{edads_pp}.

From our results, we cannot be sure that erosive processes were strong enough to erase a large portion of craters all over the surface and for all sizes. \cite{Shevchenko2008} claim that the observed chaotic regime in the motion of the Prometheus--Pandora system can play an essential role in the long-term orbital evolution of the system. In fact, both satellites were discovered by the Voyager mission and their orbital parameters were calculated. However, 15 years later when the Hubble Space Telescope observed them, their mean longitudes had shifted by some $20^{\circ}$ from the predicted values \citep{Bosh1996}. Therefore, the orbital chaos of this system may indicate that these satellites have not always been in their current orbits or otherwise may have formed more recently. If this is the case, according to our model their ages are $\sim2$ Gyr for Pandora and $\sim1$ Gyr for Prometheus. 

\subsection{Co-orbital satellites}

Janus and Epimetheus have heavily cratered surfaces that show different levels of erosion and ponding material. Additionally, Epimetheus has grooves extending between 5 and 20 km in length, possibly generated due to tensile stress related to rotational librations.

\begin{figure}[h!]
        \centerline{\scalebox{0.7}
                {\includegraphics{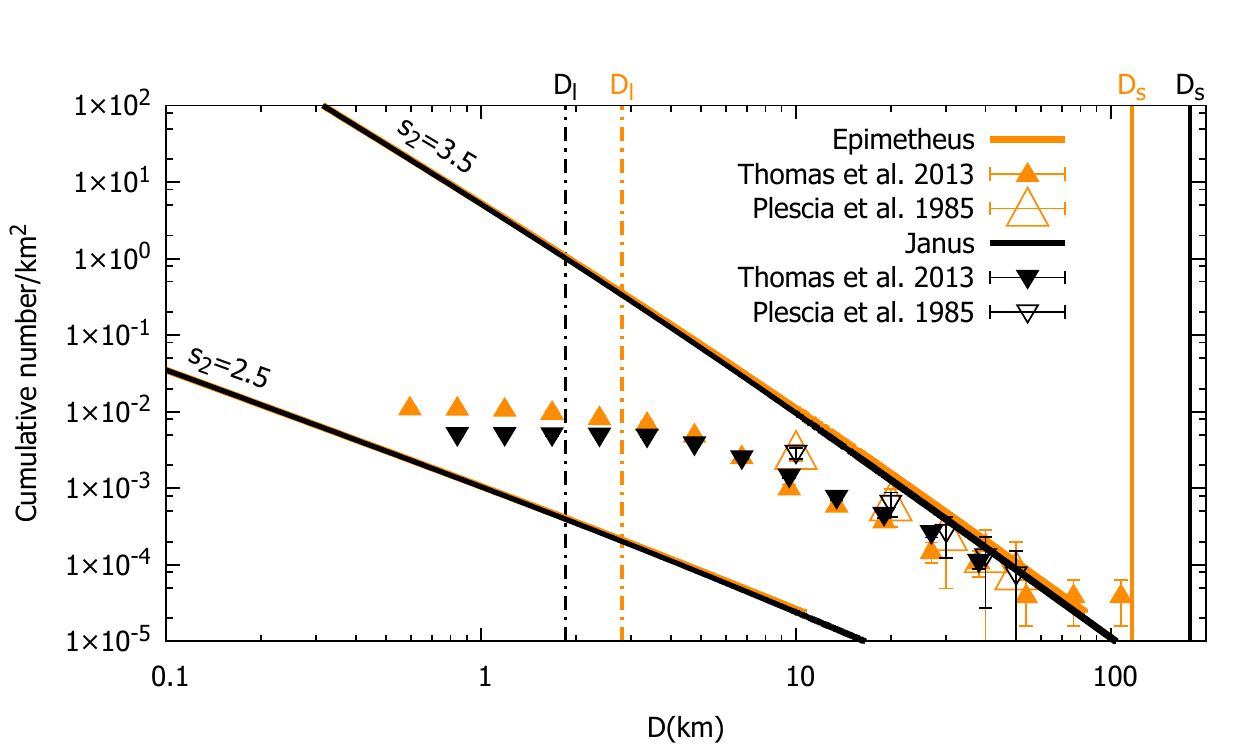}}}
        \caption{Cumulative number of craters per square kilometer as a function of crater diameter $D$ for Janus and Epimetheus. Solid lines refer to our model calculated for both values of the impactor size distribution index (top lines are for $s_2$=3.5 and bottom lines for $s_2$=2.5) and color points represent the observations presented in Plescia \& Boyce (1985) and Thomas et al. (2013). Dashed vertical
lines indicate the strength-to-gravity transition crater diameter $D_l$ on each satellite and solid vertical
lines indicate the mean body diameter $D_s$ of  each satellite.}
        \label{Coorbitals}
\end{figure}   

Figure \ref{Coorbitals} shows our results for cratering on these satellites for both values of the $s_2$ index and the crater counts for partial areas presented in \citet{Thomas2013} and \citet{Plescia1985}. The
resolution of Voyager images (between 3 and 6 km per line pair) allowed for cratering calculations of relatively large-sized craters. \citet{Plescia1983} presented crater counts based on those images for partial areas of Janus (studied area of 12361 km$^2$) and Epimetheus (studied area of 5830 km$^2$) and suggested that these satellites are fragments of a once larger body that fragmented near or just after the heavy bombardment.

The   high-resolution observations by Cassini broadened the range of detectable crater sizes, as seen in crater counts presented in \citet{Thomas2013}. The authors considered areas of 61,600 km$^2$ for Janus and of 26,393 km$^2$ for Epimetheus.     

Comparing our model to the observational crater counts, the most similar fit appears to correspond to the index $s_2$=3.5 for large craters. Our model predicts that the largest craters are 103 km in diameter (Janus) and 82 km in diameter (Epimetheus), while the largest observed craters in \citet{Thomas2013} are 38 km in diameter (Janus) and 107 km in diameter (Epimetheus). In \citet{Plescia1985} the largest observed craters are 50 km in diameter for both satellites, but those crater counts have larger errors.  We consider that the largest observed crater on  Epimetheus could have formed during the LHB phase, which is not contemplated in our model. As for the smaller craters ($D \lesssim 4$ km), we observe a deviation between our model for $s_2$=3.5 and the observed curve. This behavior is similar to what was noted for Prometheus and Pandora. 

\begin{figure}[h!]
        \centerline{\scalebox{0.7}
                {\includegraphics{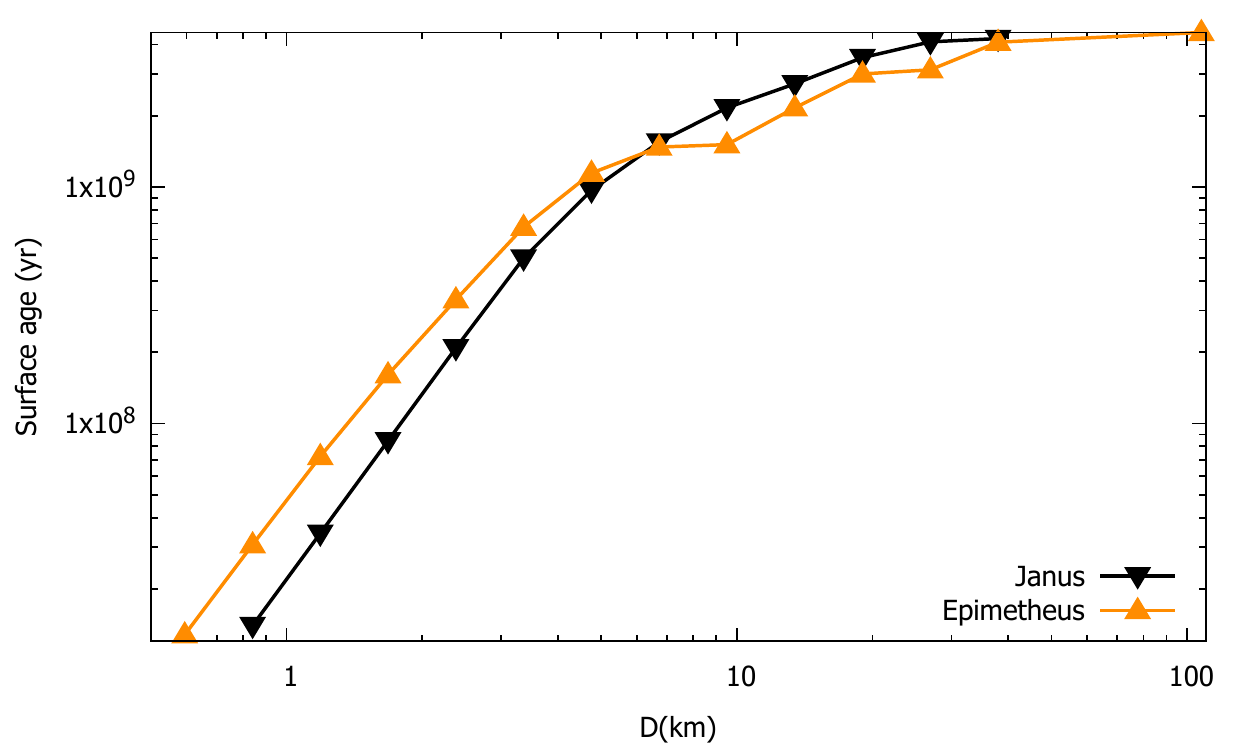}}}
        \caption{Surface age as a function of crater diameter $D$ for co-orbital satellites Janus and Epimetheus.}
        \label{edads_je}
\end{figure}

These co-orbital satellites are not directly related to the rings, but some loose material in crater floors was observed on Epimetheus \citep{Morrison2009} and the smooth appearance of this satellite may be due to this material coverage. In addition, tidal effects on these satellites with low densities are thought to generate grooves or fractures which also modify their surfaces. This can be observed in Fig. \ref{edads_je} where the surface ages of Janus and Epimetheus are plotted (Eq. \ref{agesup}). As can be seen, surface age is near 4.5 Gyr for larger craters, which may indicate that these satellites are primordial. However, for smaller craters surface age is smaller. 

We think that the behavior shown in Fig. \ref{edads_je} is due to a combination of erosion and crater saturation. Thus, surface ages calculated with our model in these satellites have to be considered as lower limits. 

\subsection{Embedded satellites}
\begin{figure}[h!]
        \centerline{\scalebox{0.7}
                {\includegraphics{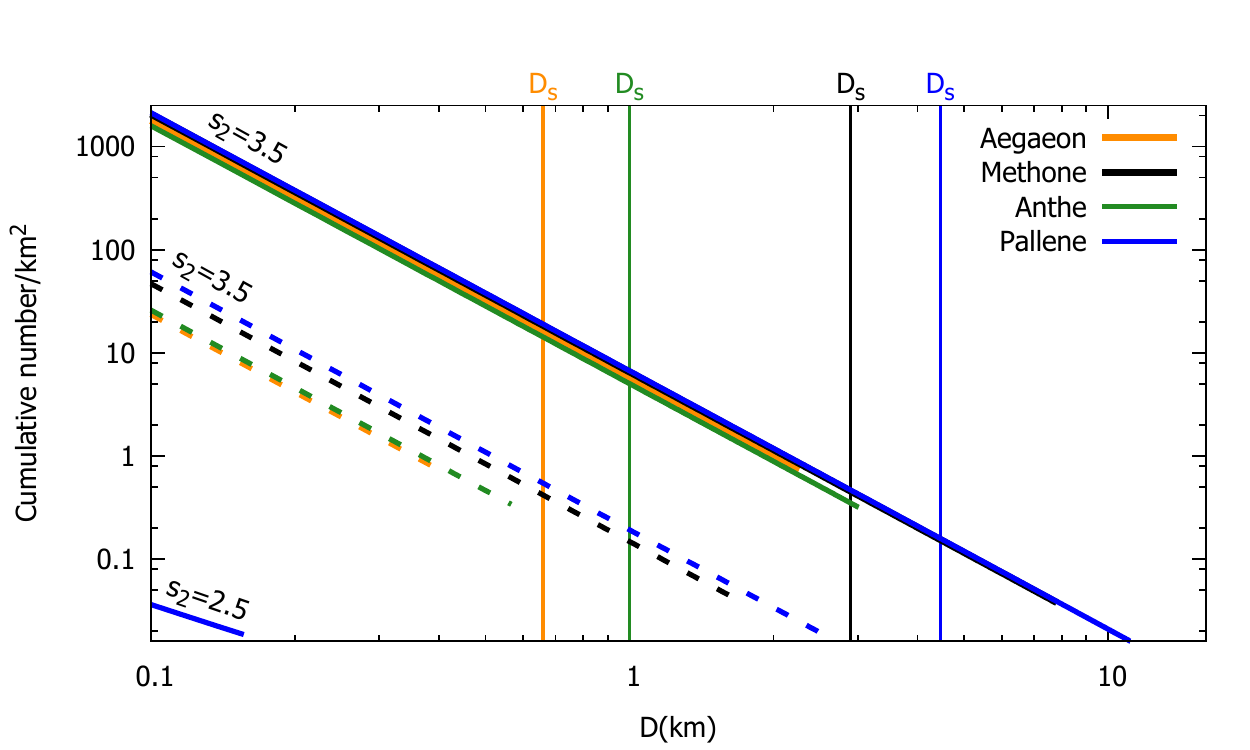}}}
        \caption{Cumulative number of craters per square kilometer as a function of crater diameter $D$ for Aegaeon, Methone, Anthe, and Pallene, calculated for both values of the impactor size distribution index $s_2$. Solid lines correspond to the cratering process considering the total age of the solar
system (top lines are for $s_2$=3.5 and bottom lines for $s_2$=2.5) and dashed lines correspond to the adapted model for the calculated age of the satellite. Vertical lines indicate the mean body diameter $D_s$ for  each satellite.}
        \label{embedded}
\end{figure}

The four satellites associated to arcs or rings orbit embedded in a stream of debris that they possibly feed through ejecta generated by impacts. Due to their small sizes, these satellites were only recently discovered by the Cassini mission, which obtained various images of their surfaces. According to these observations, no craters are visible in any of the surfaces of these satellites \citep{Thomas2013}. Methone, for example, shows a smooth surface which suggests that a mechanism that fluidizes regolith  is acting and this could probably be due to cratering processes and ejecta generation in porous objects \citep{Thomas2013}.

Our model predicts that all of these satellites suffer catastrophic collisions and consequently have formation ages ranging between 0.1 and 0.4 Gyr. In fact, if we consider that these small moons were catastrophically disrupted and reaccreted over the solar
system age (as suggested by \cite{Smith1982}), we estimate that the number of catastrophic collisions over 4.5 Gyr of evolution was: $\sim$135 for Aegaeon, $\sim$80 for Anthe, $\sim$60 for Methone, and $\sim$40 for Pallene. 

Considering only the last disruption, we can ``reset'' our calculations to the age of the last reaccretion of the satellite. Therefore, our adapted model that considers the cratering process only during the age of each satellite differs significantly from the one that considers the total age of the solar
system. Our adapted calculations of the largest craters corresponding to the $s_2$=3.5 index of the distribution are: 0.4 km (Aegaeon), 0.56 km (Anthe), 1.7 km (Methone), and 2.69 km (Pallene).
The theoretical number of craters in both cases is shown in Fig. \ref{embedded}. 

\subsection{Trojans of Tethys}
\begin{figure}[h!]
        \centerline{\scalebox{0.7}
                {\includegraphics{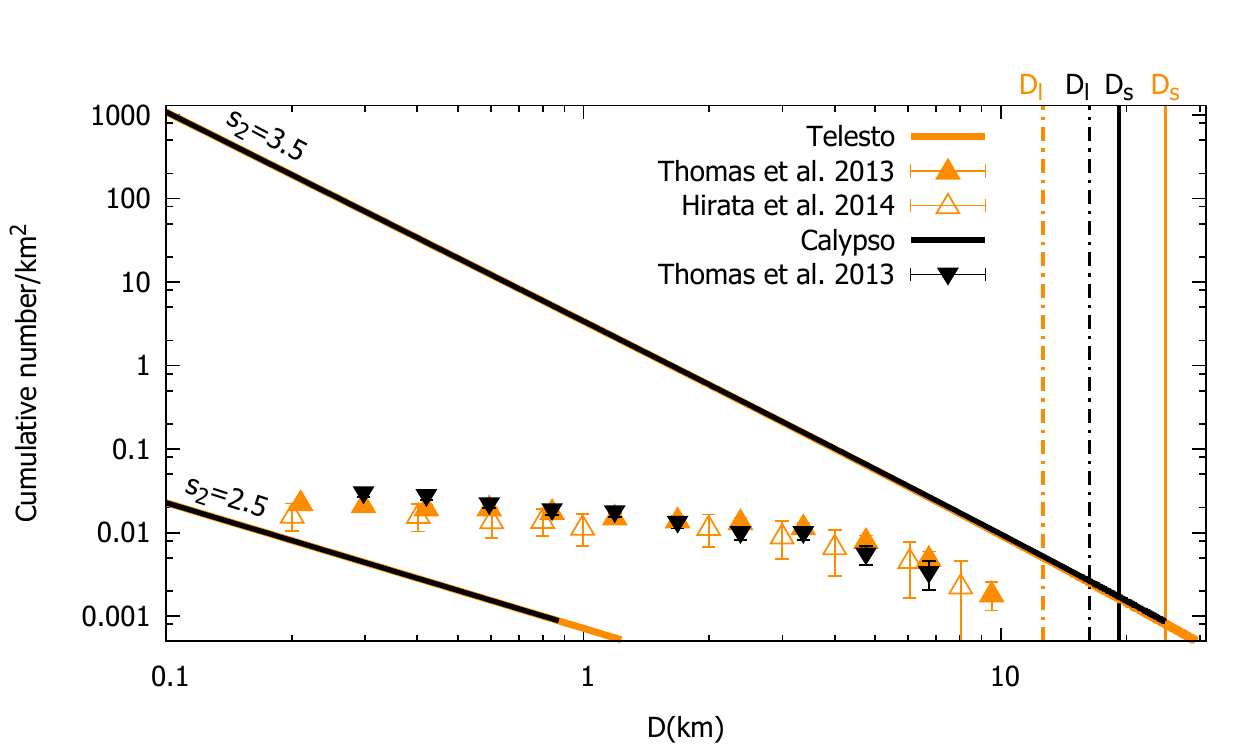}}}
        \caption{Cumulative number of craters per square kilometer as a function of crater diameter $D$ for Telesto and Calypso. Solid lines refer to our model calculated for both values of the impactor size distribution index (top lines are for $s_2$=3.5 and bottom lines for $s_2$=2.5) and color points represent the observations presented in \cite{Thomas2013} and \cite{Hirata2014}. Dashed vertical
lines indicate the strength-to-gravity transition crater diameter $D_l$ on each satellite and solid vertical
lines indicate the mean body diameter $D_s$ of  each satellite.}
        \label{tethystr}
\end{figure}

\citet{Thomas2013} describe the  surfaces of Telesto and Calypso as heavily cratered, with the presence of drainage basins and ponding of material in their craters. Particularly in Telesto, observations made by Cassini show that the crater filling is not simply rim material that has fallen in, but is instead composed of fine particles compatible with E-ring material or even ejecta coming from Tethys or Dione.

The theoretical number of craters on these Trojans is shown in Fig. \ref{tethystr} together with the observed crater counts. \citet{Thomas2013} presented crater counts calculated over areas of 1600 km$^2$ for Telesto and 910 km$^2$ for Calypso. As in the case of Prometheus and Pandora, the observed crater size distribution is between both theoretical curves.  Therefore, assuming
that Centaurs are the main source of craters on these satellites, we can  discard our model for the $s_2 = 2.5$ index. However, our model for $s_2=3.5$ predicts a greater number of craters than the observed ones for all values of $D$. In this case, the largest craters are 24 km (Calypso) and 29 km (Telesto) in diameter, while \cite{Thomas2013} presented maximum craters of 6.72 km (Calypso) and 9.51 km (Telesto) in diameter and \citet{Hirata2014} lists the largest crater observed in Telesto as 8.02 km in diameter.

Although our calculations predict that Calypso suffers one catastrophic disruption, its estimated age is 4.45 Gyr, which does not represent a large difference in the adapted crater distribution when compared to the one calculated considering the total age of the solar
system. For this reason we only include its original crater distribution in our results.

It seems possible that for the trojans of Tethys, E-ring particle showers contribute in the process of crater erosion, in some cases even erasing them completely \citep{Hirata2014}. This could also account for the more pronounced deviation of the observed distribution for smaller craters ($D \lesssim 4$ km) compared to the one corresponding to the $s_2$=3.5 index. However, a depletion of large craters does not seem probable only based on E-ring erosion. Therefore, a more accurate fit between our model and observations would be achieved if the  trojans of Tethys were captured or formed more recently. The time of formation or capture was estimated by the calculation of the surface age, $\tau_s$ (Sect. \ref{surfaceage}), since it represents the time span from the present on which craters were produced on the satellite with the theoretical cratering rate of our model. Surface ages for these satellites as a function of diameter can be seen in Fig.\ref{edad_troyanos}. Thus, Telesto and Calypso may have been captured or formed $\sim 2.8$ Gyr ago.

\subsection{Trojans of Dione}
\subsection*{Polydeuces}
\begin{figure}[!h]
        \centerline{\scalebox{0.7}
                {\includegraphics{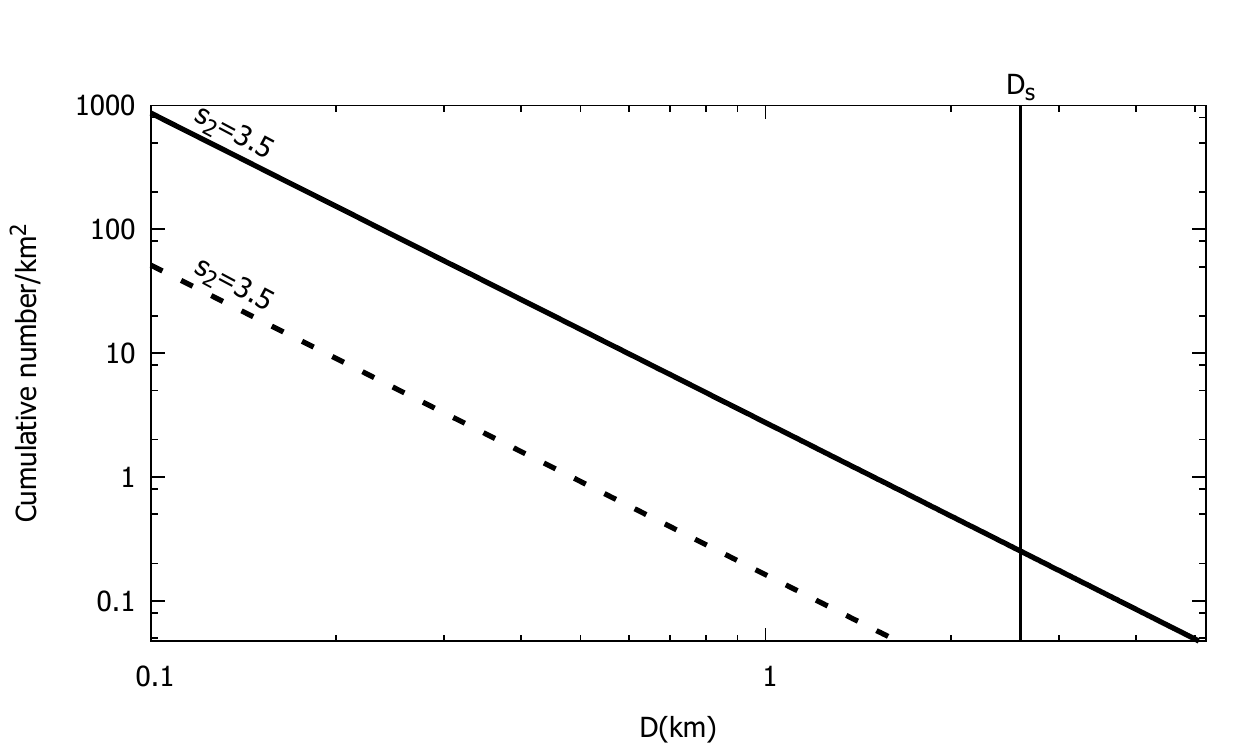}}}
        \caption{Cumulative number of craters per square kilometer as a function of crater diameter $D$ for Polydeuces. The solid line corresponds to the cratering process for the $s_2$=3.5 index considering the total age of the solar
system and the dashed line corresponds to the adapted model (for the $s_2$=3.5 index) for the calculated age of the satellite. The vertical line indicates the mean body diameter $D_S$.}
        \label{poly}
\end{figure}

Polydeuces is the smallest trojan satellite and was discovered by the Cassini mission in 2004. The model results on the crater size distribution are shown in Fig. \ref{poly}. Our calculations predict that due to its small size, it suffered various disruptions, the last one being $\sim$1 Gyr ago. In addition, our $s_2$=2.5 model does not produce any craters larger than 100 m, while if the $s_2$=3.5 index is considered the largest crater produced in the last 1 Gyr is 1.63 km in diameter. Its small size and its proximity to the E ring probably add to a smooth surface with no distinguishable craters. 

\subsection*{Helene}
\begin{figure}[h!]
        \centerline{\scalebox{0.7}
                {\includegraphics{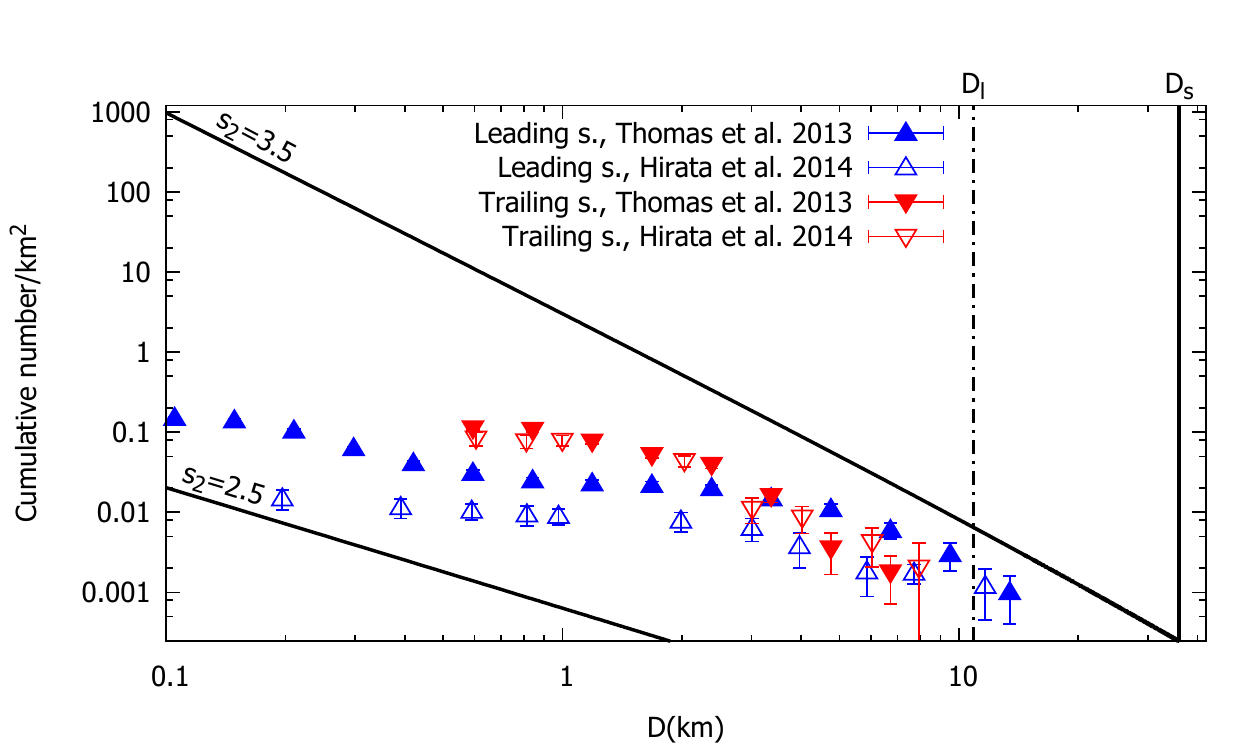}}}
        \caption{Cumulative number of craters per square kilometer as a function of crater diameter $D$ for Helene. Solid lines refer to our model calculated for both values of the impactor size distribution index (top line is for $s_2$=3.5 and bottom line for $s_2$=2.5) and color points represent the observations presented in Thomas et al. (2013) and Hirata et al. (2014). The dashed vertical
line indicates the strength-to-gravity transition crater diameter $D_l$ and the solid vertical
line indicates its mean body diameter $D_s$.} 
        \label{helenelead}
\end{figure}

Our results for crater density on Helene produced by Centaur objects are shown in Fig. \ref{helenelead}. Color  points  correspond  to  crater counts on Helene's leading and trailing sides obtained from Cassini images \citep{Thomas2013,Hirata2014}. Crater counts presented in \citet{Thomas2013} correspond to partial areas of Helene of 559 km$^2$ for its trailing side and 1004 km$^2$ for its leading side.

Helene was one of the most observed satellites by the Cassini Mission. \citet{Thomas2013} and \citet{Hirata2014} describe the large difference in crater density between different regions and conclude that it cannot be explained by a single crater population or episode \citep{Thomas2013}. \citet{Hirata2014} relate this remarkable feature to the deposition over the leading side of fine particles originated in the E ring. The sets of data presented in both papers allowed us to compare both sides with our calculations, from which we observe a better agreement between our model for $s_2$=3.5 and the trailing side, especially for larger  craters ($D > 3$ km). A possible explanation for this is that our calculations do not consider crater erosion and consequently result in a better fit for regions with well preserved (or less eroded) craters, which in the case of Helene corresponds to the trailing side.   
Our results for $s_2$=3.5 predict that the largest crater is 36 km in diameter, while observations show craters of 7.93 km \citep{Hirata2014} and 6.72 km \citep{Thomas2013} in diameter for the trailing side and 11.67 km \citep{Hirata2014} and 13.45 km \citep{Thomas2013} in diameter for the leading side.

The difference in crater density for both sides is more noticeable for craters smaller than 2 km in diameter, where erosion plays a dominant role in the surface morphology. 

\begin{figure}[h!]
        \centerline{\scalebox{0.7}
                {\includegraphics{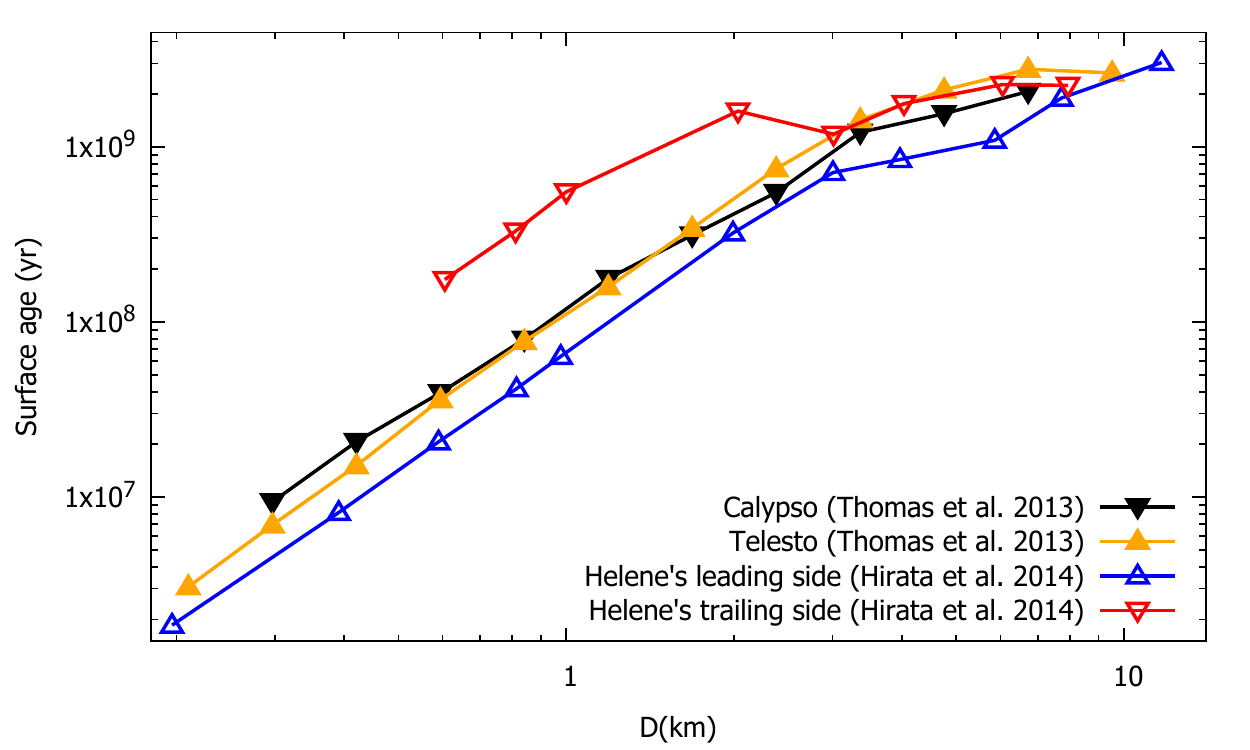}}}
        \caption{Surface age as a function of crater diameter $D$ for the trojan satellites Calypso, Telesto, and Helene.}
        \label{edad_troyanos}
\end{figure}

The surface age of Helene is shown in Fig. \ref{edad_troyanos} for both sides. The trailing side surface is older than the leading side surface, possibly due to erosion processes as mentioned. However, surface age is less than 4.5 Gyr for all the diameter values, which may evidence a more recent formation/capture time. Our results show that Helene may have been on its current orbit for the last 3.5 Gyr. 

\subsection {Hyperion}
\begin{figure}[!h]
        \centerline{\scalebox{0.7}
                {\includegraphics{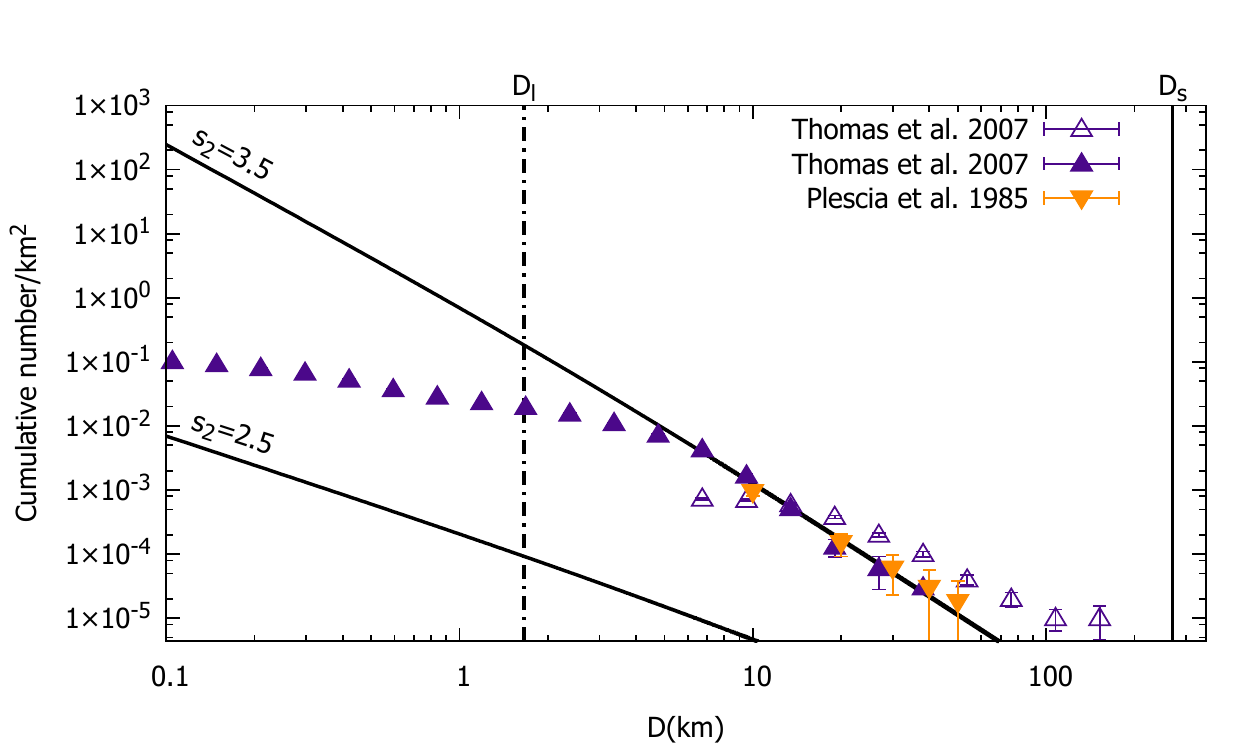}}}
        \caption{Cumulative number of craters per square kilometer as a function of crater diameter $D$ for Hyperion. Solid lines refer to our model calculated for both values of the impactor size distribution index (top line is for $s_2$=3.5 and bottom line is for $s_2$=2.5) and color points represent the observations presented in Plescia and Boyce (1985) and Thomas et al. (2007), where two different areas of the surface of the satellite were studied. The dashed vertical
line indicates the strength-to-gravity transition crater diameter $D_l$ and the solid vertical
line indicates its mean body diameter $D_s$.}  
        \label{hyperionobs}
\end{figure}

Our results for the cratering on Hyperion produced by Centaur objects are shown in Fig.~\ref{hyperionobs}. Color points correspond to crater counts on partial areas of Hyperion from Voyager \citep{Plescia1985} and Cassini images \citep{Thomas2007}. Resolution of Voyager images (9 km per line pair) only allowed for counts of large craters. \citet{Plescia1985} determined the number of craters on a counting area equal to 44,731 km$^2$ and observed a craterized surface with several 20-50 km craters and one crater with a diameter greater than 100 km.

High-resolution observations obtained from the Cassini mission allowed for the determination of the crater-size distribution up to smaller sizes. \citet{Thomas2007} show crater counts on two areas of Hyperion; 
one of them has 22,000 km$^2$ allowing for the counting of relatively large craters, the two largest being $\sim 150$ km in diameter, and the other area has 31,000 km$^2$ in which reliable crater counts reach diameters up to $\sim 1$ km.
As can be seen from Fig.~\ref{hyperionobs}, the observations nearly fit our results for $s_2 = 3.5$. Therefore, assuming that Centaurs are the main source of craters in the Saturn system, the size distribution of craters on Hyperion seems to proceed from impactors with a differential power law index of $s_2 = 3.5$. 

For this index, our model predicts that the largest crater is 69 km in diameter, while the largest observed craters are 50 km \citep{Plescia1983}, 38 km, and 152 km \citep{Thomas2007} in diameter.
 We note that the number of observed craters greater than $\sim$20 km is somewhat greater than our theoretical calculation. Considering that the difference is not a large one, we think that one possibility is that a number of those large craters could have formed during the LHB, which we do not consider in our model.
Regarding the smaller craters ($D \lesssim 7$ km), it is notable that the slope of the crater CSD becomes flatter. We think that this deviation of the observed distribution with respect to our model for $s_2$=3.5 is due to crater saturation and not to erosion, as Hyperion seems to have well preserved craters. However, \cite{Thomas2007} suggested that this downturn in the slope could be a reflection of the impactor population. Therefore, this change in the slope could be due to a combination of factors.
As we mentioned before, the combination of porous composition and great distance to the rings seems to be the key to Hyperion's well preserved craters, and a decisive factor to achieve the most similar fit for our model.
The surface of  Hyperion is heavily cratered, and the fact that the size distribution for greater craters nearly fits with our model supports the idea of an old object \citep{Plescia1983}.
 

\section{Discussion and Conclusions}

In this work, we study the cratering of the small satellites of Saturn through a previously developed method considering the Centaurs from the SD as the main impactor population. We considered a model of the SD and the evolution through the Centaur zone developed by \cite{DiSisto2007}. From this work we used the encounter files of the simulation and the method developed by \cite{DiSisto2011} to calculate the collisions of Centaurs with the small satellites and the subsequent production of craters. We assumed a CSD of SDOs that has a break at $d = 60$ km. Given the uncertainty of the size distribution at smaller sizes, we considered two differential power-law indexes for $d < 60$ km, namely $s_2=2.5$ and $3.5,$ and we expressed our results in terms of these two values. We then compared our model with the available crater counts obtained from the Cassini and Voyager observations. This comparison enabled us to better constrain the model along with some physical and dynamical properties of those satellites. 

In general, we note that the CSD with $s_2 = 3.5$ is the one with an associated crater size distribution that is more consistent with the observed crater distributions. The CSD with $s_2 = 2.5$ does not predict any of the observed craters. Another factor of uncertainty in relation to the CSD that affects the results is the "break location" from  greater to smaller SDO sizes. We assumed this to be at $d$ = 60 km (see Eq \ref{nr}), but this value is not known exactly. In order to study how a change in the break location would affect our results, we tested a different break radius at $d = 100$ km from \citet{Fraser2014}, concluding that our results do not change significantly.
Subsequently, on the basis that Centaurs are the main source of craters on the Saturnian satellites, we can discard the CSD with the $s_2 = 2.5$ index. However, a possible planetocentric source of craters should not be ruled out, given that if planetocentric debris were to be considered an important source of craters on the satellites, our results should be reviewed.

Each small satellite has a particular environment, which we analyzed in
detail in the previous section. Nevertheless, it is worth noting that in all
the studied satellites, the slope of the theoretical  crater-size
distribution (for $s_2$=3.5) is near the observed one. However, for smaller
craters ($D \lesssim 3-8$ km), the observed distribution becomes
flatter and deviates from the theoretical one. On the one hand, the
overall comparison between the theoretical crater-size distribution and the
observed one seems to indicate that the deviation between both curves for
smaller sizes may be due to erosion of craters by ring particles and/or
crater saturation. On the other hand, evidence exists that there
may be an additional break in the SFD slopes for transneptunian objects below $\sim1 - 2$ km in
diameter. This has been inferred from cratering counts on different solar-system bodies. On  Europa, the  moon of Jupiter, there is a  relative lack of small craters
(e.g., \cite{Zahnle2003}), and the results for Pluto and Charon cratering
indicated a significantly shallower crater SFD at smaller diameters
\citep{Robbins2017}. Based on observations made by the New Horizons spacecraft, \cite{Singer19} found a break in the crater SFD slope and a deficit of smaller craters ($D \lesssim 13 $ km) on Pluto and Charon. From geological considerations, these latter authors found that this fact appears to be an intrinsic property of the impactor population, which may have a break in slope at $d \sim 1 - 2$ km.  \cite{Thomas2007} suggested that the downturn in
the slope at 10 km craters on Hyperion is ``probably a reflection of the
impactor population''. In \cite{DiSisto2013}, the authors noted that for the
mid-sized Saturnian satellites, the comparison of the calculated and
observed crater CSD for different size ranges implies that the SFD of SDOs could have a new break in the range $d 
\sim 0.2-1.4$ km. For the small 
satellites we found that this possible new break in the CSD of impactors 
could exist at $d \sim 0.04 - 0.15$ km (which corresponds to $D \sim 3-8$ km), making the slope much shallower (differential slope of $\sim -1.5$). However, the majority of the
small satellites are strongly related to the rings, some even embedded in arcs or 
rings of material which may act as an exogenous source of  erosion. The distinction between the effects of the possible endogenous or exogenous  sources of erosion on differential craterization and those generated by a possible ``break'' in the size distribution of
impactors is a complex subject that has to be analyzed in a general context.
Thus, for very small craters (D $\lesssim 3-8$ km), our results considering an impactor CSD given by
Eq. \ref{nr} need to be evaluated in the context of previous considerations.

Our results suggest that Pan, Daphnis, Atlas, Aegaeon, Methone, Anthe, Pallene, Calypso, and Polydeuces have suffered one or more catastrophic collisions over the age of the solar
system for $s_2 = 3.5,$ and therefore they are probably younger than the age of the  solar
system. Consequently, they possibly formed after the formation of the planets and regular satellites; although this implies that there should be material available from which these satellites formed (or reformed). The Saturnian ring system could provide such a source material. In fact, \cite{Charnoz2010} study the formation of the small moons beyond the Roche limit from material of the rings. However, according to our results there are young moons orbiting both outside and inside the Roche limit (although within close range). Nevertheless, this matter is a complex one and will be addressed in future works. 

For those satellites that disrupted according to our model, we have calculated their formation ages, that is the time passed since their last catastrophic disruption. We note that the formation ages depend on both the size of the satellite (the smaller, the younger) and the dynamical group they belong to, those associated to arcs being the youngest, with ages of $\sim 10^8$ years. The very small satellites, specifically Daphnis, Aegaeon, Methone, Anthe, Pallene, and Polydeuces which have diameters from hundreds of meters to a few kilometers, should have suffered several disruptions over 4.5 Gyr. If we assume that after a catastrophic disruption, there remained a dense nucleus and fragments from which the satellite reaccreted, then these satellites are primordial and our calculated age corresponds to the time of the last disruption. Nevertheless, there is another possible explanation for our results regarding the various ruptures of these small satellites. If our results suggest that a given satellite on its current orbit disrupted N times in 4.5 Gyr, then there could have been N+1 primordial satellites on this orbit, from which N broke and only one remained. This is obviously an extreme case and the answer to this problem could be a combination of both possibilities. Another aspect related to the disruption of these very small satellites is the fact that the catastrophic collision remnants could constitute a possible source of planetocentric impactors on other Saturnian satellites. However, this problem is a complex one, given that it involves not only the disruption of  satellites but also their possible subsequent re-accretion from the remaining fragments in order to be able to explain their current existence. In addition, both the dynamical and physical lifetimes of the hypothetical debris disk produced in such a disruption would need to be constrained. Therefore, the possibility that this debris disk constitutes a source of planetocentric impactors should be studied in a more comprehensive way, which is out of the scope of the present investigation. It is also worth noting that if the deviation of the observed crater distribution from the theoretical model for craters smaller than $D \lesssim 3-8$ km is due, at least partially, to a break in the impactor SFD, our estimated ages increase for some of the satellites. This could result in a significant change in the calculated ages of the very small satellites (DS $ \lesssim  8$ km) for which $d_{rup}$ is below the potential break in the SFD of the  impactors.

In addition, we calculated the surface ages of the  satellites resulting in mostly young surfaces, with the exception of Hyperion, which could be related to ongoing erosion processes on these small satellites. However, we notice that for the observed surfaces of the co-orbital satellites and the Trojan satellites, the theoretical cratering curve is higher than the observed one for all the crater sizes. This could be generated by intense erosion processes on this observed part of the surface, but the erosion of large craters is highly improbable. Another possibility is that these satellites are actually not primordial but instead were captured or formed after the formation of the solar
system. In this case, Prometheus may have been captured or formed $\sim 1$ Gyr ago and Pandora $\sim 2$ Gyr ago. Furthermore, Helene may have been in its current orbit for 3.5 Gyr and Telesto and Calypso captured or formed $\sim 2.8$ Gyr ago. As in the case of satellite ages, a potential break in the impactor SFD could modify the surface ages for small crater diameters. However, as mentioned, we need to consider both processes, namely erosion or saturation and a possible new break on the impactor SFD, to exactly constrain the results. Meanwhile, our results can be  considered as lower age limits.

The model by \cite{DiSisto2007}, from which we used the encounter files to build our cratering model and obtain our results, was made by considering the observed SDOs available at that time. When analyzing the updated SDO data obtained from the JPL Small-Body Database Search Engine, we noted that the current observations seem to follow the same trend as our 2007 debiased SDO model. Therefore, unless the number of observations changes significantly, we consider the predicted model as still appropriate in general terms. However, recent observational surveys such as the Canada--France Ecliptic Plane Survey (CFEPS) \citep{Petit11} and the Outer Solar
System Origins Survey (OSSOS) \citep{Bannister18} have discovered a large number of TNOs, going deeper into the sky and characterizing and removing observational biases. Therefore, in light of these interesting discoveries we plan to contrast our model with these updated results in the near future.

\vspace*{0.5cm}
\noindent{\bf Acknowledgments:}
The authors wish to express their gratitude to the financial support by IALP, CONICET and Agencia de Promoci\'on Cient\'{\i}fica, through the grants PIP 0436 and PICT 2014-1292. This work was partially supported by research grants from CONICET and UNLP. We also wish to thank Naoyuki Hirata for kindly sharing cratering data with us and we would like to express our gratitude to Peter C. Thomas for providing the majority of the observational data presented in this work and for valuable comments and suggestions. We acknowledge the useful comments and remarks of an anonymous referee which helped us to improve this work.

\bibliographystyle{aa}
\bibliography{biblio}
\end{document}